\documentclass[fleqn,usenatbib]{mnras}

\usepackage{newtxtext,newtxmath}
\usepackage[flushleft]{threeparttable}
\usepackage{times,epsfig,natbib,graphics,longtable,lscape,threeparttable}

\usepackage[T1]{fontenc}
\usepackage{ae,aecompl}

\usepackage{graphicx}	
\usepackage{amsmath}	
\usepackage{nccmath}
\usepackage{pdflscape}	
\usepackage{multicol}

\title[H$_0$ from SNe~II]{A 5\% measurement of the Hubble-Lema\^itre constant from Type II supernovae}

\author[de Jaeger et al.]
{T. de Jaeger$^{1}$\thanks{E-mail: dejaeger@hawaii.edu},
L.~Galbany$^{2,3}$ 
A. G. Riess$^{4,5}$,
B. E. Stahl$^{6}$,
B. J. Shappee$^{1}$,
\newauthor
A. V. Filippenko$^{6}$, 
W. Zheng$^{6}$
\\
\small
$^{1}$Institute for Astronomy, University of Hawaii, 2680 Woodlawn Drive, Honolulu, HI 96822, USA.\\
$^{2}$Institute of Space Sciences (ICE, CSIC), Campus UAB, Carrer de Can Magrans, s/n, E-08193 Barcelona, Spain.\\
$^{3}$Institut d'Estudis Espacials de Catalunya (IEEC), E-08034 Barcelona, Spain.\\
$^{4}$Space Telescope Science Institute, 3700 San Martin Drive, Baltimore, MD 21218, USA.\\
$^{5}$Department of Physics \& Astronomy, Johns Hopkins University, Baltimore, MD 21218, USA.\\
$^{6}$Department of Astronomy, University of California, Berkeley, CA 94720-3411, USA.\\
}

\date{}

\pubyear{2022}

\begin{document}
\label{firstpage}
\pagerange{\pageref{firstpage}--\pageref{lastpage}}

\maketitle

\begin{abstract}
\noindent

The most stringent local measurement of the Hubble-Lema\^itre constant from Cepheid-calibrated Type Ia supernovae (SNe~Ia) differs from the value inferred via the cosmic microwave background radiation ({\it Planck}$+\Lambda$CDM) by $\sim 5\sigma$. This so-called ``Hubble tension'' has been confirmed by other independent methods, and thus does not appear to be a possible consequence of systematic errors. Here, we continue upon our prior work of using Type II supernovae to provide another, largely-independent method to measure the Hubble-Lema\^itre constant. From 13 SNe~II with geometric, Cepheid, or tip of the red giant branch (TRGB) host-galaxy distance measurements, we derive H$_0= 75.4^{+3.8}_{-3.7}$\,km\,s$^{-1}$\,Mpc$^{-1}$ (statistical errors only), consistent with the local measurement but in disagreement by $\sim 2.0\sigma$ with the {\it Planck}$+\Lambda$CDM value. Using only Cepheids ($N=7$), we find H$_0 = 77.6^{+5.2}_{-4.8}$\,km\,s$^{-1}$\,Mpc$^{-1}$, while using only TRGB ($N=5$), we derive H$_0 = 73.1^{+5.7}_{-5.3}$\,km\,s$^{-1}$\,Mpc$^{-1}$. Via 13 variants of our dataset, we derive a systematic uncertainty estimate of 1.5\,km\,s$^{-1}$\,Mpc$^{-1}$. The median value derived from these variants differs by just 0.3\,km\,s$^{-1}$\,Mpc$^{-1}$ from that produced by our fiducial model. Because we only replace SNe~Ia with SNe~II --- and we do not find statistically significant difference between the Cepheid and TRGB H$_0$ measurements --- our work reveals no indication that SNe~Ia or Cepheids could be the sources of the ``H$_0$ tension.'' We caution, however, that our conclusions rest upon a modest calibrator sample; as this sample grows in the future, our results should be verified.

\end{abstract}

\begin{keywords}
cosmology: distance scale -- galaxies: distances and redshifts -- stars: supernovae: general
\end{keywords}


\section{Introduction}

In the century since Georges Lema\^itre \citep{lemaitre27} and Edwin Hubble \citep{hubble29} discovered that the Universe is expanding, astronomers have made significant strides in measuring its current expansion rate (known as the Hubble-Lema\^itre constant, H$_0$). Traditionally, two different approaches have been employed that leverage measurements at opposite extremes of the visible Universe.
\begin{enumerate}
\item With the \emph{distance-ladder method}, relative distances to nearby galaxies in the Hubble flow (i.e., whose motions are mainly due to the expansion of the Universe) are anchored to absolute distance measurements. It is currently comprised of three steps/rungs: (i) geometric distances like Milky Way Cepheid parallaxes from {\it Gaia} EDR3 \citep{lindegren21,riess21}, detached eclipsing binary stars in the Large Magellanic Cloud \citep{pietrzynski19}, or the Keplerian motion of masers in NGC~4258 \citep{reid19,humphreys13} are used to standardise calibrators --- e.g., Cepheids or the tip of the red giant branch (TRGB); (ii) nearby Type Ia supernovae (hereafter SNe~Ia) can be calibrated by standardised calibrators --- e.g., Cepheids \citep{riess22,riess19,dhawan18,riess18b,riess18a,burns18,riess16,riess11,freedman10,riess09,sandage06,freedman01}, TRGB \citep{dhawan22,freedman21,anand21,yuan2019,freedman19,jang17b,jang17a,madore09}, or Mira variable stars \citep{huang20,whitelock08}; and (iii) the calibration to nearby SNe~Ia is applied to SNe~Ia in the Hubble flow. Owing to a series of efforts which have allowed the scientific community to build the cosmic distance ladder over several decades, such as detached eclipsing binary stars in the Large Magellanic Cloud \citep{pietrzynski19}, {\it Gaia} parallaxes \citep{lindegren21,riess21}, Cepheids \citep{leavitt12}, tip of the red giant branch (TRGB; \citealt{lee93}), and SNe~Ia in the Hubble flow (SH0ES\footnote{``Supernovae, H$_0$ for the Equation of State of Dark Energy''; \citet{riess11}.} team), the uncertainty in the local measurement of H$_0$ has improved from $\sim 10$\% \citep{freedman01} to $\pm$ 1.4\% \citep{riess22} in the last twenty years. Using 42 SNe~Ia calibrated with Cepheids, \citet{riess22} have derived the most precise estimate of H$_0$ in the late Universe: $73.04 \pm 1.04$\,km\,s$^{-1}$\,Mpc$^{-1}$. With the same technique but using 19 SNe~Ia calibrated with TRGB, \citet{freedman21} obtained H$_0 = 69.8 \pm 0.6$ (stat) $\pm$ 1.6 (sys)\,km\,s$^{-1}$\,Mpc$^{-1}$. The difference between the TRGB and Cepheid calibrations is not yet understood (possible systematics in both methods), but it is not clear whether there is any significant difference between TRGB and Cepheid distances for SN~Ia hosts. \citet{riess21} compared the only 7 hosts in common and found no difference. Also, \citet{anand21} reanalysed the TRGB distances with different data to calibrate the zero-point in NGC~4258 and also found no significant difference with Cepheid results (H$_0 = 71.5 \pm 1.8$\,km\,s$^{-1}$\,Mpc$^{-1}$). Moreover, \citet{blakeslee21} calibrated surface brightness fluctuations with Cepheids and TRGB, obtaining the same answer for each.

\item The alternate method is based on measurements of the early Universe using the sound horizon observed from the cosmic microwave background radiation (CMB; e.g., \citealt{planck18,spergel07,bennett03,jaffe01,fixsen96}). However, unlike the distance-ladder technique, this method provides only an ``inverse'' cosmic distance ladder, calibrated at redshift $z \approx 1100$ and based on the physics of the early Universe extrapolated to $z \approx 0$. Assuming a $\Lambda$ cold dark matter ($\Lambda$CDM) cosmological model, \citet{planck18} derive a value of H$_0 = 67.4 \pm 0.5$\,km\,s$^{-1}$\,Mpc$^{-1}$. Other works add an intermediate-redshift rung to anchor SNe~Ia at $z > 0.1$ and find a consistent value \citep{macaulay19}. It is important to note that all the probes from the early Universe assume that the sound horizon calculation from the standard cosmological model is correct. For this reason, \citet{baxter21} derive H$_0$ from the CMB without using information from the sound horizon scale. Their result, H$_0 = 73.5 \pm 5.3$\,km\,s$^{-1}$\,Mpc$^{-1}$, is consistent with the local measurement but different from the \citet{planck18} value.
\end{enumerate}

The discrepancy between the two approaches, also referred to as the ``H$_0$ tension,'' has reached a 5$\sigma$ level of significance using Cepheids \citep{riess22} (though only a 1--2$\sigma$ level of significance using TRGB \citealt{freedman21,anand21}). This tension is difficult to explain by invoking systematic errors, because a multitude of independent methods have confirmed it. For example, \citet{pesce2020} derived an independent H$_0$ value of $73.9 \pm 3.0$\,km\,s$^{-1}$\,Mpc$^{-1}$ using geometric distance measurements to megamaser-hosting galaxies, and \citet{blakeslee21} obtained a value of $73.3 \pm 0.7\pm 2.4$\,km\,s$^{-1}$\,Mpc$^{-1}$ from surface brightness fluctuation distances for 63 bright early-type galaxies (see \citealt{divalentino21} for a review). To date, no solution has been found to explain the tension, but a wide variety of ideas have been proposed --- e.g., the presence of additional species of neutrinos, early dark energy, decaying dark matter, or a breakdown of the general relativity (see \citealt{divalentino21} and \citealt{riess22} for reviews).

In this work, as an independent approach to test the second and third rungs of the distance-ladder method (which rely on SNe~Ia), we use Type II supernovae (SNe~II; explosions of massive, evolved, hydrogen-envelope stars via core collapse). SNe~II display a large range of peak luminosities, but can be calibrated via theoretical \citep{voglphd,schmidt94,kirshner74} and empirical methods \citep{dejaeger20,rodriguez19a,dejaeger17a,dejaeger15b,hamuy02}. Using the former, \citet{schmidt94} obtained an H$_0$ value of $73 \pm 13$\,km\,s$^{-1}$\,Mpc$^{-1}$, while with the latter, values of $69 \pm 16$\,km\,s$^{-1}$\,Mpc$^{-1}$ (standard candle method (SCM); \citealt{olivares10}) and $\sim 71 \pm 8$\,km\,s$^{-1}$\,Mpc$^{-1}$ (photospheric magnitude method; \citealt{rodriguez19a}) have been derived. More recently, by applying a refined version of the SCM \citep{dejaeger20} and using seven objects with Cepheid or TRGB independent host-galaxy distance measurements, \citet{dejaeger20b} demonstrated that SNe~II also manifest the ``H$_0$ tension'' (albeit at a low level of significance). They found an H$_0$ value of $75.8^{+5.2}_{-4.9}$\,km\,s$^{-1}$\,Mpc$^{-1}$ (stat) value, which differs by $1.4\sigma$ from the high-redshift result \citep{planck18}. Finally, using a tailored-expanding-photosphere method \citep{vogl19,vogl20}, \citet{voglphd} obtain a value of $72.3 \pm 2.8$\,km\,s$^{-1}$\,Mpc$^{-1}$, where again the uncertainties are only statistical. It is worth noting that the tailored-expanding-photosphere method is currently limited by a small sample size (only six objects) and peculiar-velocity corrections (mean $z = 0.02$), and it is affected by the systematic uncertainties of atmosphere models \citep{vogl19,dessart05,eastman96}. However, even if this method requires multiple well-calibrated spectra in the first month after the explosion, which is observationally expensive, it is a promising technique as it does not need calibrators. With this method, one can derive absolute SN~II distances and therefore measure direct H$_0$ values without the risk of introducing systematic errors from the calibrators.

Here, as in \citet{dejaeger20b}, we use the SCM to derive precise extragalactic distances, but importantly, we nearly double the number of calibrators (from 7 to 13). This allows us to derive H$_0$ with a precision of $\sim 5$\% (statistical). Section 2 describes our methodology (data, calibrators, SCM), and we present our results in Section 3. Section 4 summarises our conclusions.

\section{Method}

\subsection{Data Sample}

In this study, we consider the same SN~II sample used by \citet{dejaeger20b}, consisting of 125 objects (89 of which are at $z > 0.01$) from the following surveys: the Lick Observatory Supernova Survey (LOSS; \citealt{filippenko01}), the Carnegie Supernova Project-I (CSP-I; \citealt{ham06}), the Sloan Digital Sky Survey-II SN Survey (SDSS-II; \citealt{frieman08}), the Supernova Legacy Survey (SNLS; \citealt{astier06}), the Subaru Hyper-Suprime Cam Survey (SSP-HSC; \citealt{aihara18a,miyazaki12}), and the Dark Energy Survey Supernova Program (DES-SN; \citealt{bernstein12}). To this sample we also add four SNe~II for which we have absolute SN host distance measurements: SN~2014bc \citep{polshaw15}, SN~2017eaw \citep{vandyk19}, SN~2018aoq (unpublished Lick/KAIT data), and SN~2020yyz (unpublished {\it Hubble Space Telescope} data and public Zwicky Transient Factory data; \citealt{bellm18}). We refer the reader to \citet{dejaeger20b} and references therein for more detailed information regarding the surveys, photometric reduction, and how the magnitudes are simultaneously corrected for Milky Way extinction, $K$-corrected, and S-corrected. Note that as in \citet{dejaeger20b}, all of the CMB redshifts ($z_{\rm CMB}$) are taken from the NASA/IPAC Extragalactic Database (NED\footnote{\url{http://ned.ipac.caltech.edu/}}). Then, to account for peculiar velocities, all are corrected (henceforth referred to as $z_{\rm corr}$) using the model of \citet{carrick15}. Finally, a residual peculiar-velocity uncertainty of 250\,km\,s$^{-1}$ is added to the total redshift uncertainty in quadrature.

\subsection{Calibrator sample}\label{txt:calibrator}

This work uses 13 SNe~II having absolute distance measurements: one with a geometric distance, seven with Cepheid-derived distances, and five from the TRGB. Among these calibrators, seven were already used (and thus described) by \citet{dejaeger20b}. We list the remaining six below.
\begin{description}

\item $\bullet$ SN~2004et and SN~2017eaw in NGC~6946: \citet{dejaeger20b} did not include these objects because they had a large Milky Way extinction, and at that time, the TRGB distance was not reliable (only a few stars). In this work, we add both objects because the colour-magnitude diagram from the Extragalactic Distance Database (EDD\footnote{\url{https://edd.ifa.hawaii.edu/}}) is now well sampled. This means that unlike in \citet{anand18}, the break in the stellar luminosity function is now sharper and therefore more reliable. To consider the large Milky Way extinction, we account for it in the distance error by adding 10\% of the extinction (0.1\,mag) in quadrature. The final distance modulus used is $\mu = 29.21 \pm 0.16$\,mag.
\item $\bullet$ SN~2008bk in NGC~7793: This SN was also removed from \citet{dejaeger20b} because its distance was obtained using ground-based observations with only 11 Cepheids. However, a TRGB measurement \citep{anand21} is now available in the EDD. The distance modulus used in this work is $\mu = 27.80 \pm 0.08$\,mag.
\item $\bullet$ SN~2014bc ($\mu = 29.387 \pm 0.0568$\,mag; \citealt{reid19}) in NGC~4258 using the Keplerian motion of masers.
\item $\bullet$ SN~2018aoq ($\mu = 31.04 \pm 0.07$\,mag; \citealt{yuan20}) in NGC~4151 using Cepheids.
\item $\bullet$ SN~2020yyz ($\mu = 31.71 \pm 0.157$\,mag; \citealt{riess22}) in NGC~0976 using Cepheids.

\end{description}

The TRGB luminosities are converted into distance moduli using a zero-point calibration of $-4.01$, which is the average of many recent measurements as compiled by \citet[see their Table 3]{Li22} and an uncertainty of 0.04\,mag. Additionally, the Cepheid distances have been revised and updated from \citet{riess22}. It is important to note that because it is not clear whether there is any significant difference between TRGB and Cepheid distances for SN~Ia hosts (see Sec.~1), here we use TRGB and Cepheid distance measurements together to increase the total number of calibrators and decrease the statistical error in H$_0$. Also, in Section \ref{txt:calibrators_method}, we show that the mean SN~II luminosity from TRGB and Cepheids is consistent (differing by $\sim 0.3\sigma$), which supports the use of both calibrators together. A summary of all the calibrators available in this work and their distances can be found in Table \ref{tab:calibrators}. 

\subsection{Empirical SN~II standardisation}\label{txt:SCM}

SNe~II are not standard candles, but they are standardisable using theoretical or empirical methods. Here, we follow the methodology of \citet{dejaeger20b} and use the SCM, which leverages the correlation between SN~II luminosity and two observables: (i) the photospheric expansion velocity, and (ii) colour. Intrinsically brighter SNe~II have more-rapidly expanding photospheres and are bluer (see Figures 7 and 8 of \citealt{dejaeger20}). Therefore, for each SN, the corrected magnitude is written as

\begin{ceqn}
\begin{align}
m_{\rm corr}=\mathrm{m}+\alpha\, \mathrm{\log_{10}}\left( \frac{v_\mathrm{H\beta}}{\bar{v}_\mathrm{H\beta}} \right) - \beta (c - \bar{c})~,
\label{m_model}
\end{align}
\end{ceqn}
where $m$ is the apparent magnitude in a given passband at 43\,d after the explosion, $c$ is the colour, $v_\mathrm{H\beta}$ is the velocity measured using H$\beta$ absorption from an optical spectrum, and the overbars are used to denote averaged quantities. The nuisance parameters $\alpha$ and $\beta$ are discussed below. For more details, we refer the reader to Equations (1), (2), and (3) of \citet{dejaeger20}.

\begin{table*}
\center
\caption{Calibrator sample.} 
\begin{tabular}{lcccc}
\hline
SN name	& Host Galaxy & $\mu$ (mag) & calibrator &references\\
\hline
SN~1999em &NGC~1637 &30.26 $\pm$ 0.09 & Cepheids &\citet{dejaeger20b} (updated from \citealt{leonard03})\\
SN~1999gi &NGC~3184 &30.64 $\pm$ 0.11 & Cepheids &\citet{dejaeger20b} (updated from \citealt{leonard02b})\\
SN~2004et &NGC~6946 &29.21 $\pm$ 0.16 & TRGB &From EDD, \citet{anand21}\\
SN~2005ay &NGC~3938 &31.72 $\pm$ 0.07 & Cepheids &\citet{riess22}\\
SN~2005cs &NGC~5194/M51 &29.62 $\pm$ 0.09 & TRGB & \citet{dejaeger20b} (updated from \citealt{mcquinn17})\\
SN~2008bk &NGC~7793 &27.80 $\pm$ 0.08 & TRGB &From EDD, \citet{anand21}\\
SN~2009ib &NGC~1559 &31.49 $\pm$ 0.06 & Cepheids &\citet{riess22}\\
SN~2012aw &NGC~3351 &29.82 $\pm$ 0.09 & Cepheids &\citet{dejaeger20b} (updated from \citet{kanbur03})\\
SN~2013ej &NGC~628/M74 &29.90 $\pm$ 0.08 & TRGB &\citet{dejaeger20b} (updated from \citet{mcquinn17})\\
SN~2014bc &NGC~4258 &29.387 $\pm$ 0.0568 & Geometric &\citet{reid19}\\
SN~2017eaw &NGC~6946 &29.21 $\pm$ 0.16 & TRGB &From EDD, \citet{anand21}\\
SN~2018aoq &NGC~4151 &31.04 $\pm$ 0.07 & Cepheids &\citet{yuan20}\\
SN~2020yyz &NGC~0976 &31.71 $\pm$ 0.15 & Cepheids &\citet{riess22}\\

\hline
\label{tab:calibrators}
\end{tabular}
\end{table*}

\subsection{H$_0$ from SNe~II}\label{txt:H0_SNII}

This section describes how H$_0$ can be derived from SNe~II using the SCM. 
As the methodology is the same as that used by \citet{dejaeger20}, only a brief description is presented here.\\ 

\noindent As defined by \citet{riess11},
\begin{ceqn}
\begin{align}
\log_{10}~{{\rm H}_0}= \frac{M_i + 5\,a_i + 25}{5},
\label{eq:H0}
\end{align}
\end{ceqn}
where $a_{i}$ is the intercept of the SN~II magnitude-redshift relation (translated to $z = 0$) measured from the Hubble-flow sample and $M_{i}$ is the absolute SN~II $i$-band magnitude (at 43\,d) derived using our calibrator sample. Therefore, the approach is to fit a joint model which combines the calibrator and Hubble-flow samples to constrain $M_i$ and to determine $a_{i}$. Simultaneously, our model evaluates how close the calibrators are to the mean absolute magnitude, and, given a value of H$_0$, how close the absolute magnitudes of the Hubble-flow SNe~II are to the mean absolute magnitude.

However, as the SNe~II are not standard candles, we also need to standardise their apparent magnitudes by deriving $\alpha$ and $\beta$ from Equation \ref{m_model}. Our model thus has five free parameters: $\alpha$, $\beta$, H$_0$, $M_i$, and $\sigma_{\rm int}$, where $\sigma_{\rm int}$ is the usual uncertainty added to account for unmodelled, intrinsic SN~II scatter. As in \citet{dejaeger20}, we use the Python package \textsc{EMCEE} developed by \citet{foreman13} with 300 walkers, 2000 steps, and with uniform priors for $\alpha$, $\beta \neq 0$, H$_0 > 0$, and $M_i < 0$, and scale-free for $\sigma_{\rm int} > 0$ with $p(\sigma_{\rm int})=1/\sigma_{\rm int}$.

\section{Results}\label{txt:results}

\subsection{Calibrators}\label{txt:calibrators_method}

\begin{figure}
	\includegraphics[width=1.0\columnwidth]{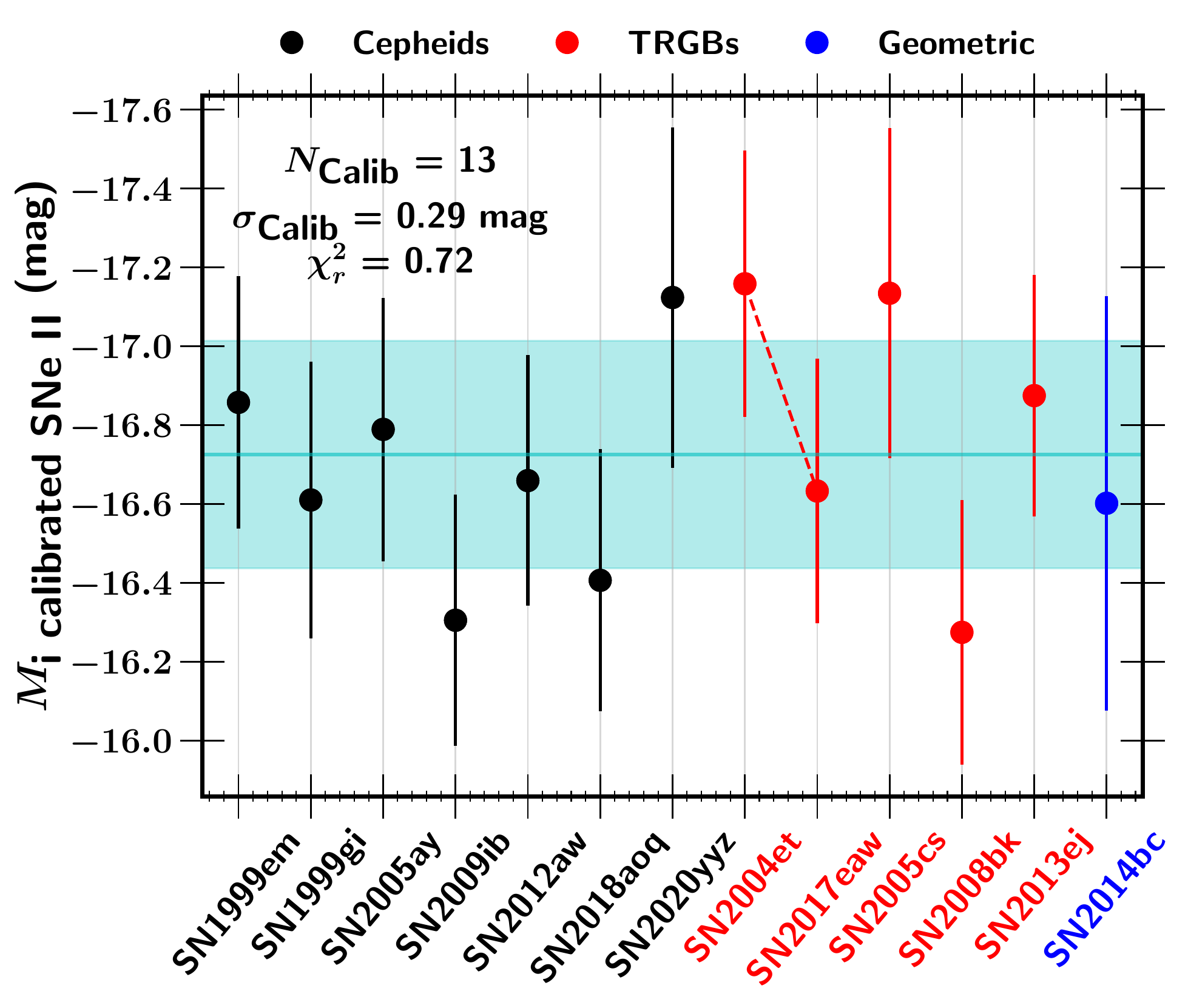}
\caption{Absolute $i$-band magnitude 43\,d after the explosion for the 13 calibrators based on Cepheid (black), TRGB (red), or geometric (blue) distances. We also present the standard deviation obtained after applying the SCM, represent it by the cyan filled region. A dashed line connecting SN~2004et and SN~2017eaw has been plotted to indicate that they are located in the same host galaxy. Note that the uncertainties include the intrinsic scatter ($\sigma_{\rm int} = 0.29$\,mag) as well as the reduced $\chi_{r}^{2}$.}
\label{fig:Mabs_cal}
\end{figure}

Following \citet{dejaeger20}, who demonstrated that the best passband to minimise the intrinsic dispersion among SNe~II in the Hubble diagram is the $i$ band, we use the same band and show, in Figure \ref{fig:Mabs_cal}, the absolute magnitudes of all 13 calibrators. The calibrators have a weighted average absolute magnitude of $-16.71$\,mag, with a dispersion of $\sigma_{\rm cal}= 0.29$\,mag --- similar to those obtained by \citet{dejaeger20} ($-16.69$\,mag and 0.24\,mag, respectively) and as expected, larger in scatter than that obtained using SNe~Ia and 42 calibrators (0.13\,mag; \citealt{riess22}). Although the method to standardise SNe~II is not as strong as the one used for SNe~Ia, the dispersion increases to 0.80\,mag when the SCM is not applied, demonstrating its utility.

It is interesting to compare the average absolute magnitude obtained for both types of calibrators. For our 5 TRGBs, we find an average absolute magnitude of $-16.81 \pm 0.33$\,mag, while for our 7 Cepheids $-16.68 \pm 0.25$\,mag. The absolute magnitude for the TRGB is slightly larger than, but fully consistent with, the Cepheids. Small-number statistics may explain the difference, as in \citet{dejaeger20} the difference was $\sim 1\sigma$ with two TRGBs and five Cepheids, while in this work it is just $\sim 0.3\sigma$.

\subsection{Hubble-Lema\^itre constant}\label{txt:fiducial}

\begin{figure*}
\centering
\includegraphics[width=2.0\columnwidth]{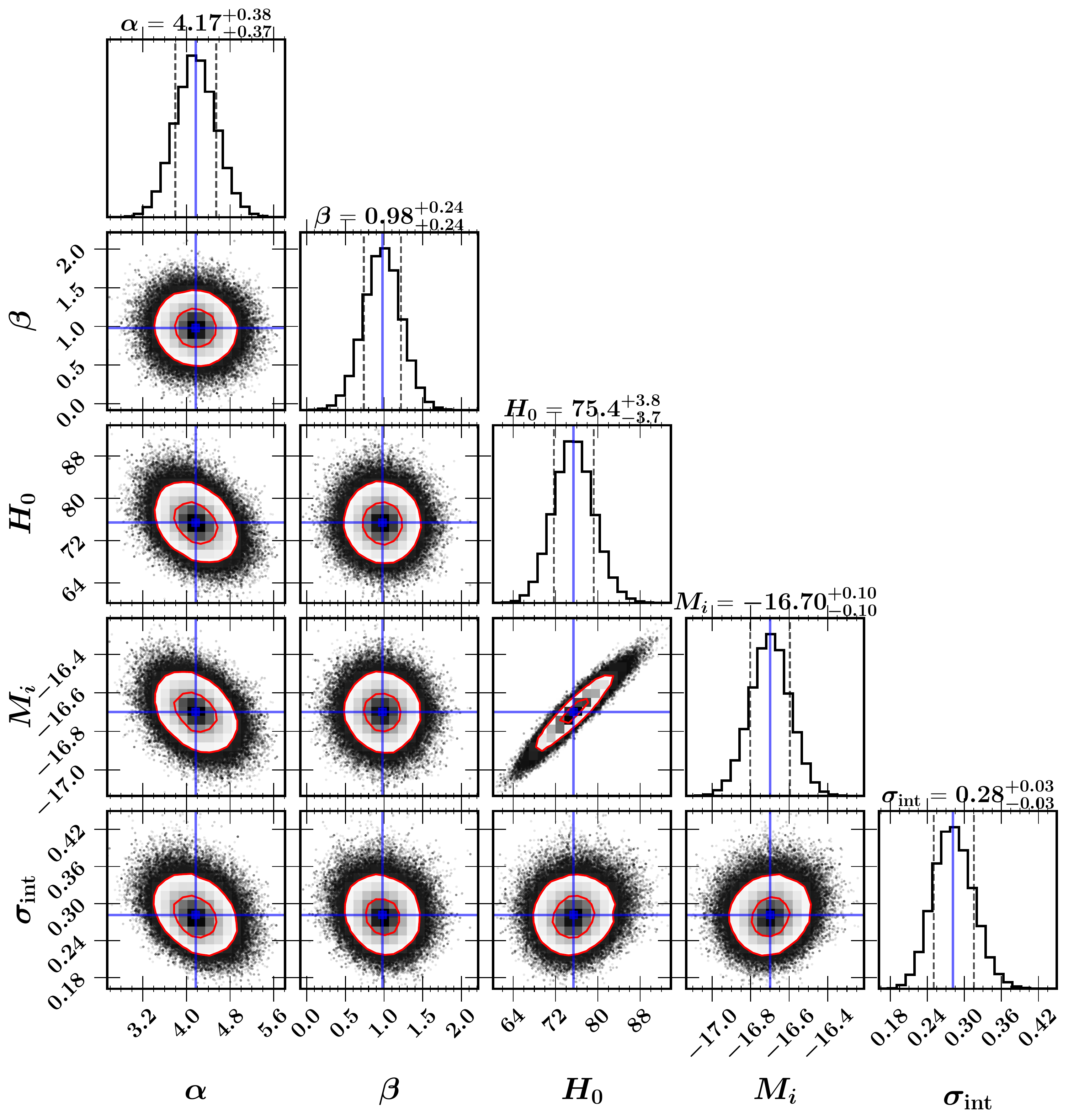}
\caption{Corner plot showing all one- and two-dimensional projections of our fitted parameters: $\alpha$, $\beta$, H$_0$, $M_i$, and $\sigma_{\rm int}$. Data points shown in grey and red contours are given at 1$\sigma$ and 2$\sigma$ (which corresponds in two dimensions to the 39\% and 86\% of the volume). For each parameter, the median value and the 16th and 84th percentile differences are shown.}
\label{fig:corner_plot_H0}
\end{figure*}

To minimise the effect of peculiar velocities we select only SNe~II with $z_{\rm corr} > 0.01$ in our Hubble-flow sample ($N=89$). With the 13 calibrators described in Section~\ref{txt:calibrator}, we obtain a median value of H$_0 = 75.4^{+3.8}_{-3.7}$\,km\,s$^{-1}$\,Mpc$^{-1}$, where the quoted uncertainties are statistical only. This value is consistent with the one derived by \citet{dejaeger20b} with seven calibrators (H$_0 = 75.8^{+5.2}_{-4.9}$\,km\,s$^{-1}$\,Mpc$^{-1}$); however, with the addition of six calibrators, we reduce the statistical uncertainty by 25\% (5.0\% vs. 6.7\%; see \citealt{dejaeger20b}). As expected and seen in Figure \ref{fig:corner_plot_H0}, the other free-fitting parameters ($\alpha$, $\beta$, $M_i$, and $\sigma_{\rm int}$) are only slightly different with respect to \citet{dejaeger20b}, as we use the same Hubble-flow sample and add six new nearby objects. Note that the intrinsic scatter derived for the SNe~II in the Hubble flow and the nearby SNe~II is consistent (0.28\,mag vs. 0.29\,mag).

Regarding the ``H$_0$ tension,'' our result is consistent with the local measurement from SNe~Ia ($73.04 \pm 1.04$\,km\,s$^{-1}$\,Mpc$^{-1}$; \citealt{riess22}), and shows a discrepancy of $2.2\sigma$ with the early-Universe value (H$_0 = 67.4 \pm 0.5$\,km\,s$^{-1}$\,Mpc$^{-1}$; \citealt{planck18}). If we use only the Cepheids to measure H$_0$ ($N=7$), we obtain H$_0 = 77.6^{+5.2}_{-4.8}$\,km\,s$^{-1}$\,Mpc$^{-1}$, while using only TRGB ($N=5$), we find H$_0 = 73.1^{+5.7}_{-5.3}$\,km\,s$^{-1}$\,Mpc$^{-1}$. There is no meaningful difference between our results derived from TRGB or from Cepheids.

\begin{figure*}
	\includegraphics[width=2.0\columnwidth]{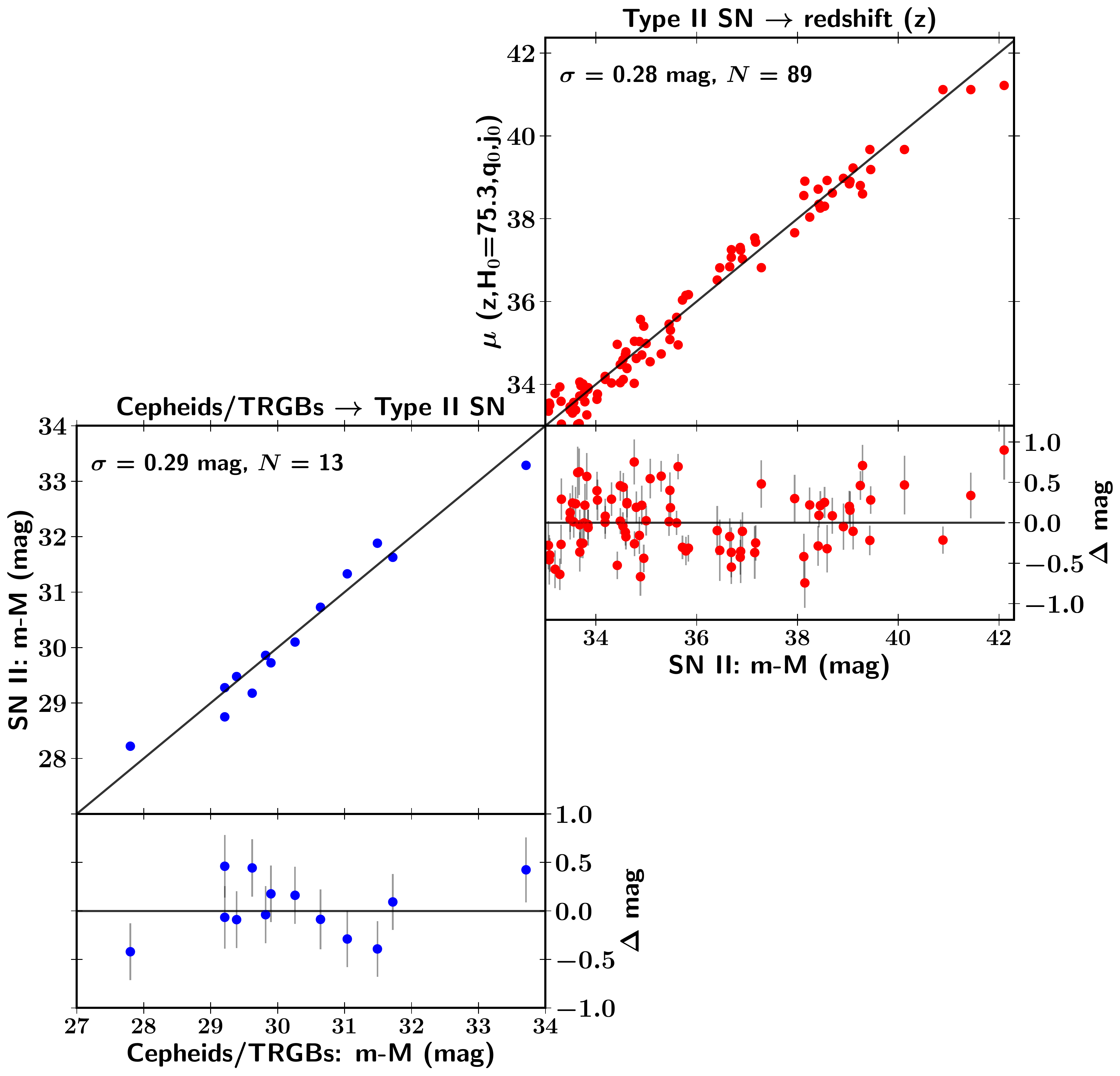}
\caption{Figure similar to the SNe~Ia figure of \citet{riess22}, representing the last two rungs of the distance ladder: Cepheid- and SN-based (bottom left), and SN- and redshift-based (top right). Blue dots represents the SNe~II with geometric, Cepheid, or TRGB distances to estimate $M_i$. Red dots are the SNe~II in the Hubble flow used to derive H$_0$.}
\label{fig:distance_ladder}
\end{figure*}

A summary of our data, H$_0$ fit, and residuals is shown in Figure \ref{fig:distance_ladder}, where we see only the second and third rungs of the distance-ladder method that have been tested in this work. The second rung allows us to calibrate and derive the SN~II absolute $i$-band magnitude using 13 calibrators (geometric, Cepheids, TRGB), while the third rung uses SNe~II in the Hubble flow to constrain H$_0$. 

\subsection{Systematic uncertainties}\label{txt:samples}

In this section, we investigate possible sources of systematic errors in our measurement. For this, we look at the effect of different cuts and calibrators on H$_0$. We summarise all the results in Table \ref{tab:sys_H0}.

First, because peculiar velocities can systematically affect H$_0$ measurements \citep{boruah21,sedgwick21}, we investigate what changes in the associated uncertainty in the recession velocities have on our determination of H$_0$. We find that changing the error to 150\,km\,s$^{-1}$ instead of 250\,km\,s$^{-1}$ only changes the value by 0.2\% ($75.3^{+4.0}_{-3.7}$\,km\,s$^{-1}$\,Mpc$^{-1}$). Then, we investigate what changes if we cut our Hubble-flow sample at $z_{\rm corr} > 0.023$ \citep{riess22}. With this cut, our Hubble-flow sample decreases to 47 SNe~II and we find a value of $77.6^{+4.7}_{-4.5}$\,km\,s$^{-1}$\,Mpc$^{-1}$ --- an increase of 2.9\% with respect to our fiducial model. If we apply a less-restrictive redshift cut and use all the SNe~II ($z_{\rm corr} > 0.0$), a decrease of 1.3\% is seen ($H_{0}=74.4^{+3.7}_{-3.3}$\,km\,s$^{-1}$\,Mpc$^{-1}$). Finally, we investigate what changes if we use uncorrected CMB-frame redshifts rather than redshifts corrected for peculiar velocities. In this case, H$_0$ decreases by 0.5\% to $75.0^{+3.8}_{-3.6}$\,km\,s$^{-1}$\,Mpc$^{-1}$. The effect on H$_0$ seen when applying different redshift cuts can be explained by peculiar velocities that are not perfectly corrected or by small-number statistics of the Hubble-flow sample (the largest difference is seen when the sample is reduced to 47 objects). 

Second, we investigate the effect of the calibrators on H$_0$. Using only Cepheids or TRBGs as calibrators causes the largest differences relative to the fiducial model. We find a difference of 2.9\% ($77.6^{+5.2}_{-4.8}$\,km\,s$^{-1}$\,Mpc$^{-1}$) and 3\% ($73.1^{+5.7}_{-5.6}$\,km\,s$^{-1}$\,Mpc$^{-1}$) with only Cepheids and only TRGBs, respectively. The small discrepancy between the TRGB and Cepheid values could hint that there might be a systematic difference between the TRGB and Cepheid methods, as possibly seen with SNe~Ia (see \citealt{riess21}; \citealt{freedman21}; \citealt{anand21}). However, our TRGB and Cepheid values are consistent, differing by $<1.0\sigma$. Also, both values are in the range of other local measures \citep{divalentino21} and statistically inconsistent with the {\it Planck}$+\Lambda$CDM value, suggesting that neither points to the source of the tension.

Finally, two SNe~II (SN~2004et and SN~2017eaw) with TRGB distance measurements have a large Milky Way extinction. If we remove them from our calibrator sample, H$_0$ increases to $77.0^{+4.4}_{-4.3}$\,km\,s$^{-1}$\,Mpc$^{-1}$ (difference of 2.2\%). We expect to find a higher value than in our fiducial model because after removing two TRGB distance measurements, the Cepheid calibrator sample size represents $\sim 63\%$ (vs. $\sim 53\%$) of all the calibrators. As the Cepheid H$_0$ value is larger than the TRGB H$_0$ value, our H$_0$ value excluding those two SNe~II from the TRGB sample will move toward a higher value than our fiducial model.

Finally, we investigate the effect of the different surveys. Using only the CSP-I sample or removing it only affects our fiducial H$_0$ measurement by 0.5\%. The major differences are seen when only the low-$z$ KAIT sample is used or removed, producing a difference of 3.0\% and 1.2\%, respectively. The largest difference could be explained by a small number of SNe~II in the Hubble flow (19) or by intrinsic SN~II differences. However, no significant differences are seen in the magnitude, velocity, and colour distributions of the CSP-I and KAIT surveys. Finally, excluding the two low-$z$ samples (CSP-I and KAIT) increases the H$_0$ value to $77.2^{+4.8}_{-4.4}$\,km\,s$^{-1}$\,Mpc$^{-1}$, a difference of 2.4\%.

All 13 H$_0$ measurements from the aforementioned analysis variants are consistent with our fiducial model. The median and standard deviation of all the variants are $75.1 \pm 1.5 $\,km\,s$^{-1}$\,Mpc$^{-1}$, which corresponds to only 0.3\,km\,s$^{-1}$\,Mpc$^{-1}$ lower than our fiducial value (only $\sim 8$\% of the statistical uncertainty). Following the conservative approach of \citet{riess19}, our systematic uncertainty is calculated as the standard deviation of our variants. From the 13 variants presented in Table \ref{tab:sys_H0}, we obtain a systematic uncertainty of $\sim 1.5$\,km\,s$^{-1}$\,Mpc$^{-1}$ ($\sim$ 2\%). Including both statistical and systematic uncertainties, our H$_0$ value is $75.4^{+3.8~{\rm(stat)}}_{-3.7~{\rm (stat)}} \pm 1.5~{\rm (sys)}$\,km\,s$^{-1}$\,Mpc$^{-1}$. This is the most precise H$_0$ value obtained from SNe~II with the SCM. Taking into account both sources of uncertainties, our value differs by $2.0\sigma$ from the high-redshift results \citep{planck18} and by only 0.6$\sigma$ from the local measurement \citep{riess22}.

\begin{table*}
\scriptsize
\caption{Free-parameter values for different sample choices.}
\begin{threeparttable}
\begin{tabular}{cccccccccccc}
\hline
Sample & Cali &$N_{\rm cali}$ &$\sigma_{\rm cali}$ &$N_{\rm SNe}$ &$\alpha$ &$\beta$ & H$_0$ &$M_i$ &$-5\,a_i$ & $\sigma_{\rm int}$ & $\Delta{{\rm H}_{0}}$\\
 & & &(mag) & & & &(km\,s$^{-1}$\,Mpc$^{-1}$) &(mag) &(mag) & (mag) &\\
\hline
\hline
Fiducial &C$+$T$+$G &13 &0.29 &89 &4.17 $^{+0.38}_{-0.37}$ &0.98 $^{+0.24}_{-0.24}$ &75.4 $^{+3.8}_{-3.7}$ &$-$16.70 $^{+0.10}_{-0.10}$ &$-$1.09 $^{+0.04}_{-0.04}$ &0.28 $^{+0.03}_{-0.03}$ & $\cdots$\\
\hline
\multicolumn{12}{c}{Peculiar-Velocity Variants} \\[-0.015cm]
\hline
$v_{\rm pec} = 150$ &C$+$T$+$G &13 &0.29 &89 &4.15 $^{+0.38}_{-0.36}$ &0.98 $^{+0.24}_{-0.25}$ &75.3 $^{+4.0}_{-3.7}$ &$-$16.70 $^{+0.10}_{-0.10}$ &$-$1.08 $^{+0.04}_{-0.04}$ &0.29 $^{+0.03}_{-0.03}$ &0.2\%\\
$z_{\rm cmb}$ &C$+$T$+$G &13 &0.29 &89 &4.11 $^{+0.37}_{-0.37}$ &1.04 $^{+0.25}_{-0.24}$ &75.0 $^{+3.8}_{-3.6}$ &$-$16.70 $^{+0.10}_{-0.10}$ &$-$1.08 $^{+0.04}_{-0.04}$ &0.28 $^{+0.03}_{-0.03}$ &0.5\% \\
$z_{\rm corr}$ $>$ 0.023 &C$+$T$+$G &13 &0.29 &47 &4.36 $^{+0.53}_{-0.51}$ &0.57 $^{+0.35}_{-0.34}$ &77.6 $^{+4.7}_{-4.5}$ &$-$16.80 $^{+0.12}_{-0.12}$ &$-$1.25 $^{+0.05}_{-0.05}$ &0.28 $^{+0.04}_{-0.04}$ &2.9\% \\
$z_{\rm corr}$ $>$ 0.0 &C$+$T$+$G &13 &0.29 &116 &4.24 $^{+0.34}_{-0.34}$ &1.08 $^{+0.23}_{-0.23}$ &74.4 $^{+3.7}_{-3.4}$ &$-$16.64 $^{+0.10}_{-0.10}$ &$-$1.00 $^{+0.03}_{-0.03}$ &0.27 $^{+0.03}_{-0.03}$ &1.3\%\\
\hline
\multicolumn{12}{c}{Calibrator Sample Variants} \\[-0.015cm]
\hline
$z_{\rm corr}$ > 0.01 &C 	&7 &0.24 &89 &4.12 $^{+0.44}_{-0.43}$ &0.88 $^{+0.25}_{-0.24}$ &77.6 $^{+5.2}_{-4.8}$ &$-$16.64 $^{+0.13}_{-0.13}$ &$-$1.09 $^{+0.04}_{-0.04}$ &0.28 $^{+0.03}_{-0.03}$ &2.9\% \\
$z_{\rm corr}$ > 0.01 &T 	&5 &0.33 &89 &4.07 $^{+0.40}_{-0.40}$ &1.04 $^{+0.28}_{-0.28}$ &73.1 $^{+5.7}_{-5.3}$ &$-$16.77 $^{+0.16}_{-0.16}$ &$-$1.09 $^{+0.04}_{-0.04}$ &0.29 $^{+0.03}_{-0.03}$ &3.0\% \\
$-$04et, 17eaw &C$+$T$+$G &11 &0.28 &89 &4.11 $^{+0.39}_{-0.38}$ &0.92 $^{+0.24}_{-0.25}$ &77.0 $^{+4.4}_{-4.3}$ &$-$16.65 $^{+0.12}_{-0.12}$ &$-$1.09 $^{+0.04}_{-0.04}$ &0.28 $^{+0.03}_{-0.03}$ &2.2\% \\
\hline
\multicolumn{12}{c}{Hubble-Flow Sample Variants} \\[-0.015cm]
\hline
Only CSP-I &C$+$T$+$G 	&13 &0.29 &37 &4.20 $^{+0.48}_{-0.47}$ &0.98 $^{+0.32}_{-0.31}$ &75.1 $^{+4.1}_{-3.9}$ &$-$16.65 $^{+0.10}_{-0.10}$ &$-$1.02 $^{+0.05}_{-0.05}$ &0.27 $^{+0.04}_{-0.04}$ &0.5\% \\
No CSP-I &C$+$T$+$G 	&13 &0.29 &52 &4.33 $^{+0.49}_{-0.48}$ &0.96 $^{+0.31}_{-0.31}$ &75.0 $^{+4.4}_{-4.1}$ &$-$16.75 $^{+0.11}_{-0.11}$ &$-$1.13 $^{+0.05}_{-0.05}$ &0.28 $^{+0.05}_{-0.05}$ &0.5\% \\
Only KAIT &C$+$T$+$G &13 &0.32 &19 &4.87 $^{+0.69}_{-0.67}$ &1.29 $^{+0.40}_{-0.39}$ &73.2 $^{+4.6}_{-4.5}$ &$-$16.66 $^{+0.11}_{-0.10}$ &$-$0.98 $^{+0.09}_{-0.08}$ &0.26 $^{+0.09}_{-0.13}$ &3.0\% \\
No KAIT &C$+$T$+$G &13 &0.29 &70 &4.00 $^{+0.40}_{-0.38}$ &0.83 $^{+0.26}_{-0.26}$ &76.3 $^{+4.0}_{-3.8}$ &$-$16.70 $^{+0.10}_{-0.10}$ &$-$1.11 $^{+0.04}_{-0.04}$ &0.27 $^{+0.03}_{-0.03}$ &1.2\% \\
CSP-I$+$KAIT &C$+$T$+$G &13 &0.29 &56 &4.36 $^{+0.43}_{-0.43}$ &1.14 $^{+0.28}_{-0.28}$ &74.4 $^{+4.1}_{-3.9}$ &$-$16.65 $^{+0.10}_{-0.11}$ &$-$1.01 $^{+0.05}_{-0.05}$ &0.28 $^{+0.04}_{-0.04}$ &1.3\% \\
``high-$z$'' &C$+$T$+$G &13 &0.29 &33 &4.11 $^{+0.53}_{-0.52}$ &0.68 $^{+0.35}_{-0.35}$ &77.2 $^{+4.8}_{-4.4}$ &$-$16.78 $^{+0.12}_{-0.11}$ &$-$1.22 $^{+0.06}_{-0.06}$ &0.26 $^{+0.05}_{-0.05}$ &2.4\% \\
\hline
\hline
\end{tabular}
Effect of systematic errors on the best-fitting values using the SCM and different samples. The fiducial line corresponds to the values obtained in Section \ref{txt:fiducial}, i.e., $z_{\rm corr} > 0.01$, 13 calibrators, and 89 SNe~II in the Hubble flow. We try different cuts in redshift ($z_{\rm corr}$), surveys (e.g., only/no CSP-I, only/no KAIT, only CSP-I$+$KAIT, only high-$z$), calibrators [Cepheids (C) and/or TRGBs (T) and/or geometric (G)], and also remove some calibrators (e.g., $-$04et and $-$17eaw for SN~2004et and SN~2017eaw). The median value with the 16th and 84th percentile differences for each parameter are given together with their statistical uncertainties. The last column, $\Delta{{\rm H}_{0}}$, corresponds to the percentage difference from the fiducial model. 
\label{tab:sys_H0}
\end{threeparttable}
\end{table*}

\subsection{Bootstrap simulation}\label{txt:bootstrap}

\begin{figure}
\centering
\includegraphics[width=1.0\columnwidth]{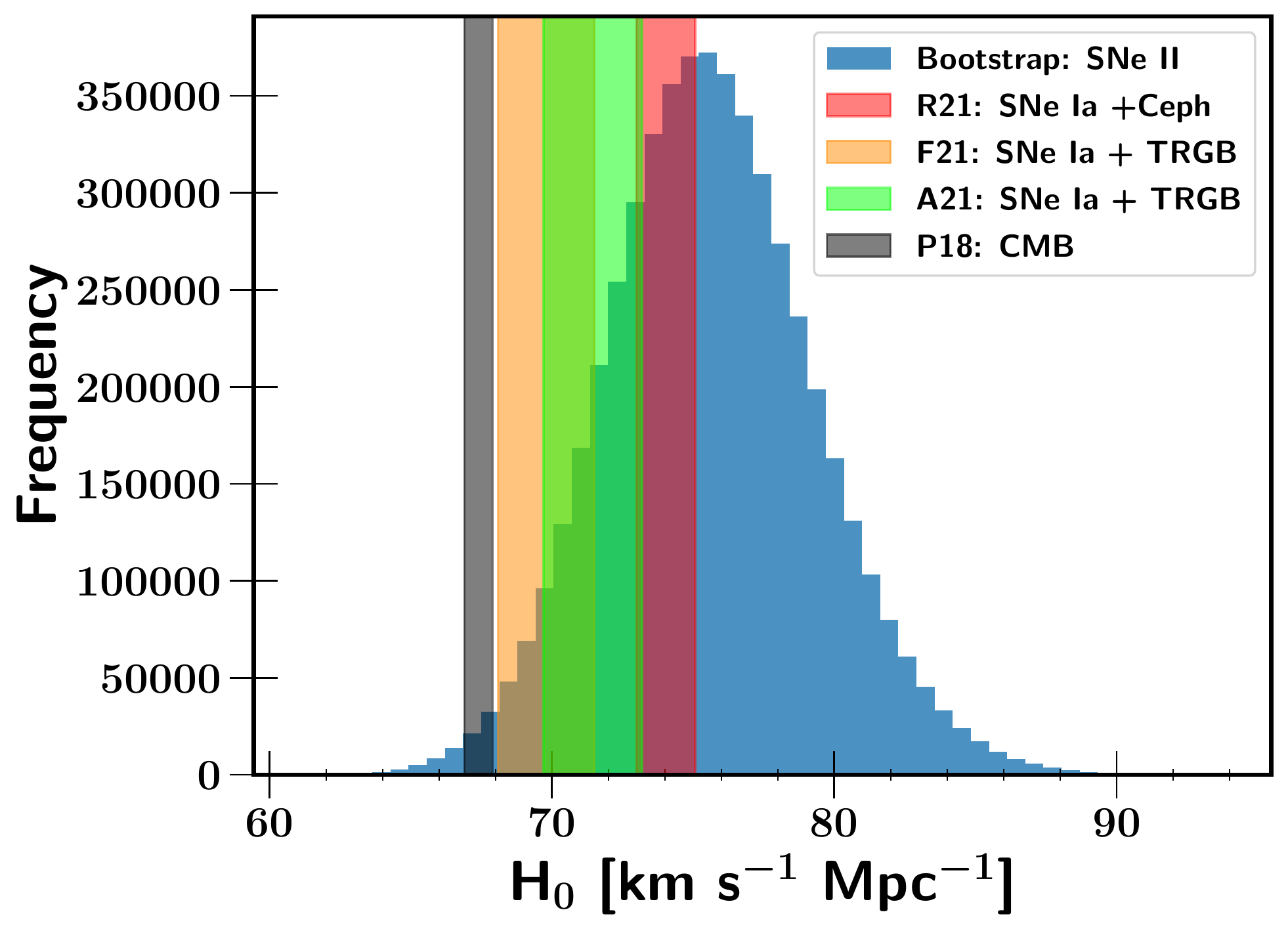}
\caption{Histogram of our bootstrap resampling of the set of calibrators, with replacement. This histogram consists of 51 bins and contains a total of 5,200,300 simulations. An average value of $75.5 \pm 3.7$\,km\,s$^{-1}$\,Mpc$^{-1}$ is derived. The red, orange, lime, and black filled regions correspond to the H$_0$ values obtained (respectively) by \citet{riess22}, \citet{freedman21}, \citet{anand21}, and \citet{planck18}. Only 1.4\% of the 5,200,300 H$_0$ values are smaller than 67.4 + 0.5\,km\,s$^{-1}$\,Mpc$^{-1}$ \citep{planck18}.}
\label{fig:bootstrap}
\end{figure}

We perform a bootstrap resampling of the set of calibrators, with replacement (see Figure~\ref{fig:bootstrap}), to study the calibrator effects on H$_0$. With 13 calibrators, we explore a total of 5,200,300 possibilities ($25!/13!12!$) and obtain a median value of $75.5 \pm 3.7$\,km\,s$^{-1}$\,Mpc$^{-1}$. The peak of the distribution is consistent with the original value and the local measurements using SNe~Ia \citep{riess22}, but almost does not overlap with the {\it Planck}$+\Lambda$CDM value. Only 1.4\% of the 5,200,300 H$_0$ samples are smaller than 67.9\,km\,s$^{-1}$\,Mpc$^{-1}$, which corresponds to {\it Planck}$+\Lambda$CDM value $+ 1\sigma$. Finally, as in \citet{dejaeger20b}, our distribution also extends to large H$_0$ values (85--95\,km\,s$^{-1}$\,Mpc$^{-1}$), and this behaviour is driven by the faintest calibrators (SN~2009ib, SN~2018aoq, and SN~2008bk).

\section{Conclusions}

In this work, we test the second and the third rungs of the SN~Ia distance ladder. For this purpose, we use SNe~II to provide an independent measurement of H$_0$. With 13 objects having geometric, Cepheid, or TRGB host-galaxy distance measurements, we derive H$_0 = 75.4^{+3.8}_{-3.7}$\,km\,s$^{-1}$\,Mpc$^{-1}$, where the quoted uncertainties are statistical only. 

By analysing 13 variants to our fiducial model, we also investigate the possible sources of systematic error. We find that all 13 H$_0$ measurements are consistent with our fiducial model, and the median value only differs by 0.3\,km\,s$^{-1}$\,Mpc$^{-1}$. From our 13 variants, we obtain a standard deviation of $\sim 1.5$\,km\,s$^{-1}$\,Mpc$^{-1}$ ($\sim 2$\%), which we interpret as an estimate of the systematic error in the SCM. Combining systematic and statistical uncertainties, we derive a value of $75.4^{+3.8~}_{-3.7}$ (stat) $\pm 1.5~{\rm (sys)}$\,km\,s$^{-1}$\,Mpc$^{-1}$. Our value is consistent with the local measurement \citep{riess22} and differs by $2.0\sigma$ from the high-redshift results \citep{planck18}. Therefore, this demonstrates that there is no evidence that SNe~Ia are the source of the ``H$_0$ tension''; the third rung of the cosmic distance ladder, yielded by SNe~Ia and~SNe II, is consistent.

We also perform a bootstrap simulation to study the calibrator effects on H$_0$. The peak of our distribution is consistent with the local measurements using SNe~Ia \citep{riess22} but almost does not overlap with the {\it Planck}$+\Lambda$CDM value. Only 1.4\% of the 5,200,300 H$_0$ values are smaller than 67.9\,km\,s$^{-1}$\,Mpc$^{-1}$ which corresponds to the {\it Planck}$+\Lambda$CDM value $+ 1\sigma$. 

Finally, with the availability of two sources of calibration, Cepheids or TRGB, we investigate the role of either in the ``H$_0$ tension.'' With seven Cepheids or five TRGB, we derive consistent values which differ by $< 1.0\sigma$ (difference of 4.5\,km\,s$^{-1}$\,Mpc$^{-1}$ between Cepheids and TRGB). Both values are also in the range of several other local measures \citep{divalentino21}. Thus, despite the larger uncertainties of our values, we find no indication of Cepheids or TRGB as the source of the ``H$_0$ tension.'' This is in good agreement with the results from \citet{blakeslee21}, \citet{kourkchi22}, \citet{anand21}, and \citet{riess22}, who found no significant difference in H$_0$ between the use of Cepheids and TRGB.

With upcoming studies, we will increase the number of SNe~II in the Hubble flow and reduce the systematic uncertainties due to peculiar velocities. Also, as shown in this paper, with a larger number of calibrators, we will be able to reduce our statistical uncertainty. Finally, having more Cepheid and TRGB distance measurements will allow us to better test the second rung of the distance ladder and see whether there is a systematic difference between both calibrators. 

\section*{Acknowledgements}

We thank the referee for their comments on the manuscript, which helped improve it. Support for T.d.J. has been provided by U.S. NSF grants AST-1908952 and AST-1911074. 
L.G. acknowledges financial support from the Spanish Ministerio de Ciencia e Innovaci\'on (MCIN), the Agencia Estatal de Investigaci\'on (AEI) 10.13039/501100011033, and the European Social Fund (ESF) ``Investing in your future" under the 2019 Ram\'on y Cajal program RYC2019-027683-I and the PID2020-115253GA-I00 HOSTFLOWS project, from Centro Superior de Investigaciones Cient\'ificas (CSIC) under the PIE project 20215AT016, and the program Unidad de Excelencia Mar\'ia de Maeztu CEX2020-001058-M.
Support for A.V.F.'s supernova research at U.C. Berkeley has been provided by the NSF through grant AST-1211916, the TABASGO Foundation, Gary and Cynthia Bengier, Marc J. Staley (whose fellowship partially funded B.E.S. whilst contributing to the work presented herein as a graduate student), the Christopher R. Redlich Fund, the Sylvia and Jim Katzman Foundation, and the Miller Institute for Basic Research in Science (A.V.F. was a Miller Senior Fellow). B.J.S. is supported by U.S. NSF grants AST-1907570, AST-1908952, AST-1920392, and AST-1911074. The work of the CSP-I has been supported by the U.S. NSF under grants AST-0306969, AST-0607438, and AST-1008343.

KAIT and its ongoing operation were made possible by donations from Sun Microsystems, Inc., the Hewlett-Packard Company, AutoScope Corporation, Lick Observatory, the U.S. NSF, the University of California, the Sylvia \& Jim Katzman Foundation, and the TABASGO Foundation. Research at Lick Observatory is partially supported by a generous gift from Google. This research used the Savio computational cluster resource provided by the Berkeley Research Computing program at U.C. Berkeley (supported by the U.C. Berkeley Chancellor, Vice Chancellor for Research, and Chief Information Officer).

This paper is based in part on data collected at the Subaru Telescope and retrieved from the HSC data archive system, which is operated by the Subaru Telescope and Astronomy Data Center at the National Astronomical Observatory of Japan (NAOJ). The Hyper Suprime-Cam (HSC) collaboration includes the astronomical communities of Japan and Taiwan, and Princeton University. The HSC instrumentation and software were developed by the NAOJ, the Kavli Institute for the Physics and Mathematics of the Universe (Kavli IPMU), the University of Tokyo, the High Energy Accelerator Research Organization (KEK), the Academia Sinica Institute for Astronomy and Astrophysics in Taiwan (ASIAA), and Princeton University. Funding was contributed by the FIRST program from the Japanese Cabinet Office, the Ministry of Education, Culture, Sports, Science and Technology (MEXT), the Japan Society for the Promotion of Science (JSPS), the Japan Science and Technology Agency (JST), the Toray Science Foundation, NAOJ, Kavli IPMU, KEK, ASIAA, and Princeton University. 

The Pan-STARRS1 Surveys (PS1) were made possible through contributions of the Institute for Astronomy, the University of Hawaii, the Pan-STARRS Project Office, the Max-Planck Society and its participating institutes, the Max Planck Institute for Astronomy, Heidelberg and the Max Planck Institute for Extraterrestrial Physics, Garching, The Johns Hopkins University, Durham University, the University of Edinburgh, Queen's University Belfast, the Harvard-Smithsonian Center for Astrophysics, the Las Cumbres Observatory Global Telescope Network Incorporated, the National Central University of Taiwan, the Space Telescope Science Institute, the National Aeronautics and Space Administration (NASA) under grant No. NNX08AR22G issued through the Planetary Science Division of the NASA Science Mission Directorate, the U.S. NSF under grant AST-1238877, the University of Maryland, and Eotvos Lorand University (ELTE). This paper makes use of software developed for the Large Synoptic Survey Telescope. We thank the LSST Project for making their code available as free software at http://dm.lsst.org. 

Some of the data presented herein were obtained at the W. M. Keck Observatory, which is operated as a scientific partnership among the California Institute of Technology, the University of California, and NASA; the observatory was made possible by the generous financial support of the W. M. Keck Foundation. This work is based in part on data produced at the Canadian Astronomy Data Centre as part of the CFHT Legacy Survey, a collaborative project of the National Research Council of Canada and the French Centre National de la Recherche Scientifique. The work is also based on observations obtained at the Gemini Observatory, which is operated by the Association of Universities for Research in Astronomy, Inc., under a cooperative agreement with the U.S. NSF on behalf of the Gemini partnership: the U.S. NSF, the STFC (United Kingdom), the National Research Council (Canada), CONICYT (Chile), the Australian Research Council (Australia), CNPq (Brazil), and CONICET (Argentina). This research used observations from Gemini program numbers GN-2005A-Q-11, GN-2005B-Q-7, GN-2006A-Q-7, GS-2005A-Q-11, GS-2005B-Q-6, and GS-2008B-Q-56. This research has made use of the NASA/IPAC Extragalactic Database (NED), which is operated by the Jet Propulsion Laboratory, California Institute of Technology, under contract with NASA, and of data provided by the Central Bureau for Astronomical Telegrams.

Funding for the DES Projects has been provided by the U.S. Department of Energy, the U.S. NSF, the Ministry of Science and Education of Spain, the Science and Technology Facilities Council of the United Kingdom, the Higher Education Funding Council for England, the National Center for Supercomputing Applications at the University of Illinois at Urbana-Champaign, the Kavli Institute of Cosmological Physics at the University of Chicago, the Center for Cosmology and Astro-Particle Physics at the Ohio State University, the Mitchell Institute for Fundamental Physics and Astronomy at Texas A\&M University, Financiadora de Estudos e Projetos, Fundacao Carlos Chagas Filho de Amparo \'a Pesquisa do Estado do Rio de Janeiro, Conselho Nacional de Desenvolvimento Cient\'ifico e Tecnol\'ogico and the Minist\'erio da Ci\^encia, Tecnologia e Inovacao, the Deutsche Forschungsgemeinschaft and the Collaborating Institutions in the Dark Energy Survey.

The DES data management system is supported by the U.S. NSF under grant AST-1138766. The DES participants from Spanish institutions are partially supported by MINECO under grants AYA2012-39559, ESP2013-48274, FPA2013-47986, and Centro de Excelencia Severo Ochoa SEV-2012-0234. Research leading to these results has received funding from the European Research Council under the European Union's Seventh Framework Programme (FP7/2007-2013) including ERC grant agreements 240672, 291329, and 306478. This research uses resources of the National Energy Research Scientific Computing Center, a DOE Office of Science User Facility supported by the Office of Science of the U.S. Department of Energy under Contract No. DE-AC02-05CH11231.\\

\textit{Software:} astropy \citep{astropy}, Matplotlib \citep{matplotlib}, Numpy \citep{numpy}, Scipy \citep{scipy}, triangle v0.1.1. Zenodo. 10.5281/zenodo.11020
 

\section*{Data Availability Statements}
The majority of the data have already been published and can be found in \citet{poznanski09} (KAIT-P09), \citet{andrea10} (SDSS-SN), \citet{dejaeger17b} (HSC), \citet{dejaeger19} (KAIT-d19), and \citet{dejaeger20} (DES-SN).
CSP-I and SNLS data will be shared on reasonable request to the corresponding author.



\bsp	
\label{lastpage}

\begin{thebibliography}{}
\makeatletter
\relax
\def\mn@urlcharsother{\let\do\@makeother \do\$\do\&\do\#\do\^\do\_\do\%\do\~}
\def\mn@doi{\begingroup\mn@urlcharsother \@ifnextchar [ {\mn@doi@}
  {\mn@doi@[]}}
\def\mn@doi@[#1]#2{\def\@tempa{#1}\ifx\@tempa\@empty \href
  {http://dx.doi.org/#2} {doi:#2}\else \href {http://dx.doi.org/#2} {#1}\fi
  \endgroup}
\def\mn@eprint#1#2{\mn@eprint@#1:#2::\@nil}
\def\mn@eprint@arXiv#1{\href {http://arxiv.org/abs/#1} {{\tt arXiv:#1}}}
\def\mn@eprint@dblp#1{\href {http://dblp.uni-trier.de/rec/bibtex/#1.xml}
  {dblp:#1}}
\def\mn@eprint@#1:#2:#3:#4\@nil{\def\@tempa {#1}\def\@tempb {#2}\def\@tempc
  {#3}\ifx \@tempc \@empty \let \@tempc \@tempb \let \@tempb \@tempa \fi \ifx
  \@tempb \@empty \def\@tempb {arXiv}\fi \@ifundefined
  {mn@eprint@\@tempb}{\@tempb:\@tempc}{\expandafter \expandafter \csname
  mn@eprint@\@tempb\endcsname \expandafter{\@tempc}}}

\bibitem[\protect\citeauthoryear{{Aihara} et~al.,}{{Aihara}
  et~al.}{2018}]{aihara18a}
{Aihara} H.,  et~al., 2018, \mn@doi [\pasj] {10.1093/pasj/psx066}, \href
  {https://ui.adsabs.harvard.edu/abs/2018PASJ...70S...4A} {70, S4}

\bibitem[\protect\citeauthoryear{{Anand}, {Rizzi}  \& {Tully}}{{Anand}
  et~al.}{2018}]{anand18}
{Anand} G.~S.,  {Rizzi} L.,   {Tully} R.~B.,  2018, \mn@doi [\aj]
  {10.3847/1538-3881/aad3b2}, \href
  {https://ui.adsabs.harvard.edu/abs/2018AJ....156..105A} {156, 105}

\bibitem[\protect\citeauthoryear{{Anand}, {Tully}, {Rizzi}, {Riess}  \&
  {Yuan}}{{Anand} et~al.}{2021}]{anand21}
{Anand} G.~S.,  {Tully} R.~B.,  {Rizzi} L.,  {Riess} A.~G.,   {Yuan} W.,  2021,
  arXiv e-prints, \href {https://ui.adsabs.harvard.edu/abs/2021arXiv210800007A}
  {p. arXiv:2108.00007}

\bibitem[\protect\citeauthoryear{{Astier}, {Guy}, {Regnault}  et~al.}{{Astier}
  et~al.}{2006}]{astier06}
{Astier} P.,  {Guy} J.,  {Regnault} N.,   et~al., 2006, \mn@doi [\aap]
  {10.1051/0004-6361:20054185}, \href
  {http://adsabs.harvard.edu/abs/2006A%26A...447...31A} {447, 31}

\bibitem[\protect\citeauthoryear{{Astropy Collaboration} et~al.,}{{Astropy
  Collaboration} et~al.}{2013}]{astropy}
{Astropy Collaboration} et~al., 2013, \mn@doi [\aap]
  {10.1051/0004-6361/201322068}, \href
  {https://ui.adsabs.harvard.edu/abs/2013A&A...558A..33A} {558, A33}

\bibitem[\protect\citeauthoryear{{Baxter} \& {Sherwin}}{{Baxter} \&
  {Sherwin}}{2021}]{baxter21}
{Baxter} E.~J.,  {Sherwin} B.~D.,  2021, \mn@doi [\mnras]
  {10.1093/mnras/staa3706}, \href
  {https://ui.adsabs.harvard.edu/abs/2021MNRAS.501.1823B} {501, 1823}

\bibitem[\protect\citeauthoryear{{Bellm} et~al.,}{{Bellm}
  et~al.}{2019}]{bellm18}
{Bellm} E.~C.,  et~al., 2019, \mn@doi [\pasp] {10.1088/1538-3873/aaecbe}, \href
  {https://ui.adsabs.harvard.edu/abs/2019PASP..131a8002B} {131, 018002}

\bibitem[\protect\citeauthoryear{{Bennett}, {Hill}, {Hinshaw}
  et~al.}{{Bennett} et~al.}{2003}]{bennett03}
{Bennett} C.~L.,  {Hill} R.~S.,  {Hinshaw} G.,   et~al., 2003, \mn@doi [\apjs]
  {10.1086/377253}, \href {http://adsabs.harvard.edu/abs/2003ApJS..148....1B}
  {148, 1}

\bibitem[\protect\citeauthoryear{{Bernstein}, {Kessler}, {Kuhlmann}
  et~al.}{{Bernstein} et~al.}{2012}]{bernstein12}
{Bernstein} J.~P.,  {Kessler} R.,  {Kuhlmann} S.,   et~al., 2012, \mn@doi
  [\apj] {10.1088/0004-637X/753/2/152}, \href
  {http://adsabs.harvard.edu/abs/2012ApJ...753..152B} {753, 152}

\bibitem[\protect\citeauthoryear{{Blakeslee}, {Jensen}, {Ma}, {Milne}  \&
  {Greene}}{{Blakeslee} et~al.}{2021}]{blakeslee21}
{Blakeslee} J.~P.,  {Jensen} J.~B.,  {Ma} C.-P.,  {Milne} P.~A.,   {Greene}
  J.~E.,  2021, \mn@doi [\apj] {10.3847/1538-4357/abe86a}, \href
  {https://ui.adsabs.harvard.edu/abs/2021ApJ...911...65B} {911, 65}

\bibitem[\protect\citeauthoryear{{Boruah}, {Hudson}  \& {Lavaux}}{{Boruah}
  et~al.}{2021}]{boruah21}
{Boruah} S.~S.,  {Hudson} M.~J.,   {Lavaux} G.,  2021, \mn@doi [\mnras]
  {10.1093/mnras/stab2320}, \href
  {https://ui.adsabs.harvard.edu/abs/2021MNRAS.507.2697B} {507, 2697}

\bibitem[\protect\citeauthoryear{{Burns} et~al.,}{{Burns}
  et~al.}{2018}]{burns18}
{Burns} C.~R.,  et~al., 2018, \mn@doi [\apj] {10.3847/1538-4357/aae51c}, \href
  {https://ui.adsabs.harvard.edu/abs/2018ApJ...869...56B} {869, 56}

\bibitem[\protect\citeauthoryear{{Carrick}, {Turnbull}, {Lavaux}  \&
  {Hudson}}{{Carrick} et~al.}{2015}]{carrick15}
{Carrick} J.,  {Turnbull} S.~J.,  {Lavaux} G.,   {Hudson} M.~J.,  2015, \mn@doi
  [\mnras] {10.1093/mnras/stv547}, \href
  {https://ui.adsabs.harvard.edu/abs/2015MNRAS.450..317C} {450, 317}

\bibitem[\protect\citeauthoryear{{D'Andrea}, {Sako}, {Dilday}
  et~al.}{{D'Andrea} et~al.}{2010}]{andrea10}
{D'Andrea} C.~B.,  {Sako} M.,  {Dilday} B.,   et~al., 2010, \mn@doi [\apj]
  {10.1088/0004-637X/708/1/661}, \href
  {http://adsabs.harvard.edu/abs/2010ApJ...708..661D} {708, 661}

\bibitem[\protect\citeauthoryear{{Dessart} \& {Hillier}}{{Dessart} \&
  {Hillier}}{2005}]{dessart05}
{Dessart} L.,  {Hillier} D.~J.,  2005, \mn@doi [\aap]
  {10.1051/0004-6361:20053217}, \href
  {http://adsabs.harvard.edu/abs/2005A%26A...439..671D} {439, 671}

\bibitem[\protect\citeauthoryear{{Dhawan}, {Jha}  \& {Leibundgut}}{{Dhawan}
  et~al.}{2018}]{dhawan18}
{Dhawan} S.,  {Jha} S.~W.,   {Leibundgut} B.,  2018, \mn@doi [\aap]
  {10.1051/0004-6361/201731501}, \href
  {https://ui.adsabs.harvard.edu/abs/2018A&A...609A..72D} {609, A72}

\bibitem[\protect\citeauthoryear{{Dhawan} et~al.,}{{Dhawan}
  et~al.}{2022}]{dhawan22}
{Dhawan} S.,  et~al., 2022, arXiv e-prints, \href
  {https://ui.adsabs.harvard.edu/abs/2022arXiv220304241D} {p. arXiv:2203.04241}

\bibitem[\protect\citeauthoryear{{de Jaeger}, {Gonz{\'a}lez-Gait{\'a}n},
  {Anderson}  et~al.}{{de Jaeger} et~al.}{2015}]{dejaeger15b}
{de Jaeger} T.,  {Gonz{\'a}lez-Gait{\'a}n} S.,  {Anderson} J.~P.,   et~al.,
  2015, \mn@doi [\apj] {10.1088/0004-637X/815/2/121}, \href
  {http://adsabs.harvard.edu/abs/2015ApJ...815..121D} {815, 121}

\bibitem[\protect\citeauthoryear{{de Jaeger}, {Galbany}, {Filippenko}
  et~al.}{{de Jaeger} et~al.}{2017a}]{dejaeger17b}
{de Jaeger} T.,  {Galbany} L.,  {Filippenko} A.~V.,   et~al., 2017a, \mn@doi
  [\mnras] {10.1093/mnras/stx2300}, \href
  {http://adsabs.harvard.edu/abs/2017MNRAS.472.4233D} {472, 4233}

\bibitem[\protect\citeauthoryear{{de Jaeger}, {Gonz{\'a}lez-Gait{\'a}n},
  {Hamuy}  et~al.}{{de Jaeger} et~al.}{2017b}]{dejaeger17a}
{de Jaeger} T.,  {Gonz{\'a}lez-Gait{\'a}n} S.,  {Hamuy} M.,   et~al., 2017b,
  \mn@doi [\apj] {10.3847/1538-4357/835/2/166}, \href
  {http://adsabs.harvard.edu/abs/2017ApJ...835..166D} {835, 166}

\bibitem[\protect\citeauthoryear{{de Jaeger} et~al.,}{{de Jaeger}
  et~al.}{2019}]{dejaeger19}
{de Jaeger} T.,  et~al., 2019, \mn@doi [\mnras] {10.1093/mnras/stz2714}, \href
  {https://ui.adsabs.harvard.edu/abs/2019MNRAS.490.2799D} {490, 2799}

\bibitem[\protect\citeauthoryear{{de Jaeger} et~al.,}{{de Jaeger}
  et~al.}{2020a}]{dejaeger20}
{de Jaeger} T.,  et~al., 2020a, \mn@doi [\mnras] {10.1093/mnras/staa1402},
  \href {https://ui.adsabs.harvard.edu/abs/2020MNRAS.tmp.1546D} {}

\bibitem[\protect\citeauthoryear{{de Jaeger}, {Stahl}, {Zheng}, {Filippenko},
  {Riess}  \& {Galbany}}{{de Jaeger} et~al.}{2020b}]{dejaeger20b}
{de Jaeger} T.,  {Stahl} B.~E.,  {Zheng} W.,  {Filippenko} A.~V.,  {Riess}
  A.~G.,   {Galbany} L.,  2020b, \mn@doi [\mnras] {10.1093/mnras/staa1801},
  \href {https://ui.adsabs.harvard.edu/abs/2020MNRAS.496.3402D} {496, 3402}


\bibitem[\protect\citeauthoryear{{Di Valentino} et~al.,}{{Di Valentino}
  et~al.}{2021}]{divalentino21}
{Di Valentino} E.,  et~al., 2021, arXiv e-prints, \href
  {https://ui.adsabs.harvard.edu/abs/2021arXiv210301183D} {p. arXiv:2103.01183}

\bibitem[\protect\citeauthoryear{{Eastman}, {Schmidt}  \& {Kirshner}}{{Eastman}
  et~al.}{1996}]{eastman96}
{Eastman} R.~G.,  {Schmidt} B.~P.,   {Kirshner} R.,  1996, \mn@doi [\apj]
  {10.1086/177563}, \href {http://adsabs.harvard.edu/abs/1996ApJ...466..911E}
  {466, 911}

\bibitem[\protect\citeauthoryear{{Filippenko}, {Li}, {Treffers}  \&
  {Modjaz}}{{Filippenko} et~al.}{2001}]{filippenko01}
{Filippenko} A.~V.,  {Li} W.~D.,  {Treffers} R.~R.,   {Modjaz} M.,  2001, in
  {Paczynski} B.,  {Chen} W.-P.,   {Lemme} C.,  eds,  Astronomical Society of
  the Pacific Conference Series Vol. 246, IAU Colloq. 183: Small Telescope
  Astronomy on Global Scales. p.~121

\bibitem[\protect\citeauthoryear{{Fixsen}, {Cheng}, {Gales}  et~al.}{{Fixsen}
  et~al.}{1996}]{fixsen96}
{Fixsen} D.~J.,  {Cheng} E.~S.,  {Gales} J.~M.,   et~al., 1996, \mn@doi [\apj]
  {10.1086/178173}, \href {http://adsabs.harvard.edu/abs/1996ApJ...473..576F}
  {473, 576}

\bibitem[\protect\citeauthoryear{{Foreman-Mackey}, {Hogg}, {Lang}  \&
  {Goodman}}{{Foreman-Mackey} et~al.}{2013}]{foreman13}
{Foreman-Mackey} D.,  {Hogg} D.~W.,  {Lang} D.,   {Goodman} J.,  2013, \mn@doi
  [\pasp] {10.1086/670067}, \href
  {http://adsabs.harvard.edu/abs/2013PASP..125..306F} {125, 306}

\bibitem[\protect\citeauthoryear{{Freedman}}{{Freedman}}{2021}]{freedman21}
{Freedman} W.~L.,  2021, \mn@doi [\apj] {10.3847/1538-4357/ac0e95}, \href
  {https://ui.adsabs.harvard.edu/abs/2021ApJ...919...16F} {919, 16}

\bibitem[\protect\citeauthoryear{{Freedman} \& {Madore}}{{Freedman} \&
  {Madore}}{2010}]{freedman10}
{Freedman} W.~L.,  {Madore} B.~F.,  2010, \mn@doi [\araa]
  {10.1146/annurev-astro-082708-101829}, \href
  {https://ui.adsabs.harvard.edu/abs/2010ARA&A..48..673F} {48, 673}

\bibitem[\protect\citeauthoryear{{Freedman}, {Madore}, {Gibson}
  et~al.}{{Freedman} et~al.}{2001}]{freedman01}
{Freedman} W.~L.,  {Madore} B.~F.,  {Gibson} B.~K.,   et~al., 2001, \mn@doi
  [\apj] {10.1086/320638}, \href
  {http://adsabs.harvard.edu/abs/2001ApJ...553...47F} {553, 47}

\bibitem[\protect\citeauthoryear{{Freedman} et~al.,}{{Freedman}
  et~al.}{2019}]{freedman19}
{Freedman} W.~L.,  et~al., 2019, \mn@doi [\apj] {10.3847/1538-4357/ab2f73},
  \href {https://ui.adsabs.harvard.edu/abs/2019ApJ...882...34F} {882, 34}

\bibitem[\protect\citeauthoryear{{Frieman}, {Bassett}, {Becker}
  et~al.}{{Frieman} et~al.}{2008}]{frieman08}
{Frieman} J.~A.,  {Bassett} B.,  {Becker} A.,   et~al., 2008, \mn@doi [\aj]
  {10.1088/0004-6256/135/1/338}, \href
  {http://adsabs.harvard.edu/abs/2008AJ....135..338F} {135, 338}

\bibitem[\protect\citeauthoryear{{Hamuy} \& {Pinto}}{{Hamuy} \&
  {Pinto}}{2002}]{hamuy02}
{Hamuy} M.,  {Pinto} P.~A.,  2002, \mn@doi [\apjl] {10.1086/339676}, \href
  {http://adsabs.harvard.edu/abs/2002ApJ...566L..63H} {566, L63}

\bibitem[\protect\citeauthoryear{{Hamuy}, {Folatelli}, {Morrell}
  et~al.}{{Hamuy} et~al.}{2006}]{ham06}
{Hamuy} M.,  {Folatelli} G.,  {Morrell} N.~I.,   et~al., 2006, \mn@doi [\pasp]
  {10.1086/500228}, \href {http://adsabs.harvard.edu/abs/2006PASP..118....2H}
  {118, 2}

\bibitem[\protect\citeauthoryear{Harris et~al.,}{Harris et~al.}{2020}]{numpy}
Harris C.~R.,  et~al., 2020, \mn@doi [Nature] {10.1038/s41586-020-2649-2}, 585,
  357

\bibitem[\protect\citeauthoryear{{Huang} et~al.,}{{Huang}
  et~al.}{2020}]{huang20}
{Huang} C.~D.,  et~al., 2020, \mn@doi [\apj] {10.3847/1538-4357/ab5dbd}, \href
  {https://ui.adsabs.harvard.edu/abs/2020ApJ...889....5H} {889, 5}

\bibitem[\protect\citeauthoryear{{Hubble}}{{Hubble}}{1929}]{hubble29}
{Hubble} E.,  1929, \mn@doi [Proceedings of the National Academy of Science]
  {10.1073/pnas.15.3.168}, \href
  {http://adsabs.harvard.edu/abs/1929PNAS...15..168H} {15, 168}

\bibitem[\protect\citeauthoryear{{Humphreys}, {Reid}, {Moran}, {Greenhill}  \&
  {Argon}}{{Humphreys} et~al.}{2013}]{humphreys13}
{Humphreys} E.~M.~L.,  {Reid} M.~J.,  {Moran} J.~M.,  {Greenhill} L.~J.,
  {Argon} A.~L.,  2013, \mn@doi [\apj] {10.1088/0004-637X/775/1/13}, \href
  {https://ui.adsabs.harvard.edu/abs/2013ApJ...775...13H} {775, 13}

\bibitem[\protect\citeauthoryear{Hunter}{Hunter}{2007}]{matplotlib}
Hunter J.~D.,  2007, \mn@doi [Computing in Science \& Engineering]
  {10.1109/MCSE.2007.55}, 9, 90

\bibitem[\protect\citeauthoryear{{Jaffe}, {Ade}, {Balbi}  et~al.}{{Jaffe}
  et~al.}{2001}]{jaffe01}
{Jaffe} A.~H.,  {Ade} P.~A.,  {Balbi} A.,   et~al., 2001, \mn@doi [Physical
  Review Letters] {10.1103/PhysRevLett.86.3475}, \href
  {http://adsabs.harvard.edu/abs/2001PhRvL..86.3475J} {86, 3475}

\bibitem[\protect\citeauthoryear{{Jang} \& {Lee}}{{Jang} \&
  {Lee}}{2017a}]{jang17a}
{Jang} I.~S.,  {Lee} M.~G.,  2017a, \mn@doi [\apj]
  {10.3847/1538-4357/835/1/28}, \href
  {https://ui.adsabs.harvard.edu/abs/2017ApJ...835...28J} {835, 28}

\bibitem[\protect\citeauthoryear{{Jang} \& {Lee}}{{Jang} \&
  {Lee}}{2017b}]{jang17b}
{Jang} I.~S.,  {Lee} M.~G.,  2017b, \mn@doi [\apj]
  {10.3847/1538-4357/836/1/74}, \href
  {https://ui.adsabs.harvard.edu/abs/2017ApJ...836...74J} {836, 74}

\bibitem[\protect\citeauthoryear{{Kanbur}, {Ngeow}, {Nikolaev}, {Tanvir}  \&
  {Hendry}}{{Kanbur} et~al.}{2003}]{kanbur03}
{Kanbur} S.~M.,  {Ngeow} C.,  {Nikolaev} S.,  {Tanvir} N.~R.,   {Hendry} M.~A.,
   2003, \mn@doi [\aap] {10.1051/0004-6361:20031373}, \href
  {http://adsabs.harvard.edu/abs/2003A%26A...411..361K} {411, 361}

\bibitem[\protect\citeauthoryear{{Kirshner} \& {Kwan}}{{Kirshner} \&
  {Kwan}}{1974}]{kirshner74}
{Kirshner} R.~P.,  {Kwan} J.,  1974, \mn@doi [\apj] {10.1086/153123}, \href
  {http://adsabs.harvard.edu/abs/1974ApJ...193...27K} {193, 27}

\bibitem[\protect\citeauthoryear{{Kourkchi}, {Tully}, {Courtois}, {Dupuy}  \&
  {Guinet}}{{Kourkchi} et~al.}{2022}]{kourkchi22}
{Kourkchi} E.,  {Tully} R.~B.,  {Courtois} H.~M.,  {Dupuy} A.,   {Guinet} D.,
  2022, \mn@doi [\mnras] {10.1093/mnras/stac303}, \href
  {https://ui.adsabs.harvard.edu/abs/2022MNRAS.tmp..331K} {}

\bibitem[\protect\citeauthoryear{{Leavitt} \& {Pickering}}{{Leavitt} \&
  {Pickering}}{1912}]{leavitt12}
{Leavitt} H.~S.,  {Pickering} E.~C.,  1912, Harvard College Observatory
  Circular, \href {https://ui.adsabs.harvard.edu/abs/1912HarCi.173....1L} {173,
  1}

\bibitem[\protect\citeauthoryear{{Lee}, {Freedman}  \& {Madore}}{{Lee}
  et~al.}{1993}]{lee93}
{Lee} M.~G.,  {Freedman} W.~L.,   {Madore} B.~F.,  1993, \mn@doi [\apj]
  {10.1086/173334}, \href
  {https://ui.adsabs.harvard.edu/abs/1993ApJ...417..553L} {417, 553}

\bibitem[\protect\citeauthoryear{{Lema{\^i}tre}}{{Lema{\^i}tre}}{1927}]{lemaitre27}
{Lema{\^i}tre} G.,  1927, Annales de la Soci{\'e}t{\'e} Scientifique de
  Bruxelles, \href {http://adsabs.harvard.edu/abs/1927ASSB...47...49L} {47, 49}

\bibitem[\protect\citeauthoryear{{Leonard}, {Filippenko}, {Li}
  et~al.}{{Leonard} et~al.}{2002}]{leonard02b}
{Leonard} D.~C.,  {Filippenko} A.~V.,  {Li} W.,   et~al., 2002, \mn@doi [\aj]
  {10.1086/343771}, \href {http://adsabs.harvard.edu/abs/2002AJ....124.2490L}
  {124, 2490}

\bibitem[\protect\citeauthoryear{{Leonard}, {Kanbur}, {Ngeow}  \&
  {Tanvir}}{{Leonard} et~al.}{2003}]{leonard03}
{Leonard} D.~C.,  {Kanbur} S.~M.,  {Ngeow} C.~C.,   {Tanvir} N.~R.,  2003,
  \mn@doi [\apj] {10.1086/376831}, \href
  {http://adsabs.harvard.edu/abs/2003ApJ...594..247L} {594, 247}

\bibitem[\protect\citeauthoryear{{Li}, {Casertano}  \& {Riess}}{{Li}
  et~al.}{2022}]{Li22}
{Li} S.,  {Casertano} S.,   {Riess} A.~G.,  2022, arXiv e-prints, \href
  {https://ui.adsabs.harvard.edu/abs/2022arXiv220211110L} {p. arXiv:2202.11110}

\bibitem[\protect\citeauthoryear{{Lindegren} et~al.,}{{Lindegren}
  et~al.}{2021}]{lindegren21}
{Lindegren} L.,  et~al., 2021, \mn@doi [\aap] {10.1051/0004-6361/202039709},
  \href {https://ui.adsabs.harvard.edu/abs/2021A&A...649A...2L} {649, A2}

\bibitem[\protect\citeauthoryear{{Macaulay} et~al.,}{{Macaulay}
  et~al.}{2019}]{macaulay19}
{Macaulay} E.,  et~al., 2019, \mn@doi [\mnras] {10.1093/mnras/stz978}, \href
  {https://ui.adsabs.harvard.edu/abs/2019MNRAS.486.2184M} {486, 2184}

\bibitem[\protect\citeauthoryear{{Madore}, {Mager}  \& {Freedman}}{{Madore}
  et~al.}{2009}]{madore09}
{Madore} B.~F.,  {Mager} V.,   {Freedman} W.~L.,  2009, \mn@doi [\apj]
  {10.1088/0004-637X/690/1/389}, \href
  {https://ui.adsabs.harvard.edu/abs/2009ApJ...690..389M} {690, 389}

\bibitem[\protect\citeauthoryear{{McQuinn}, {Skillman}, {Dolphin}, {Berg}  \&
  {Kennicutt}}{{McQuinn} et~al.}{2017}]{mcquinn17}
{McQuinn} K. B.~W.,  {Skillman} E.~D.,  {Dolphin} A.~E.,  {Berg} D.,
  {Kennicutt} R.,  2017, \mn@doi [\aj] {10.3847/1538-3881/aa7aad}, \href
  {https://ui.adsabs.harvard.edu/abs/2017AJ....154...51M} {154, 51}

\bibitem[\protect\citeauthoryear{{Miyazaki} et~al.,}{{Miyazaki}
  et~al.}{2012}]{miyazaki12}
{Miyazaki} S.,  et~al., 2012, \mn@doi [Proc. SPIE] {10.1117/12.926844}, \href
  {https://ui.adsabs.harvard.edu/abs/2012SPIE.8446E..0ZM} {8446, 84460Z}

\bibitem[\protect\citeauthoryear{{Olivares E.}, {Hamuy}, {Pignata}
  et~al.}{{Olivares E.} et~al.}{2010}]{olivares10}
{Olivares E.} F.,  {Hamuy} M.,  {Pignata} G.,   et~al., 2010, \mn@doi [\apj]
  {10.1088/0004-637X/715/2/833}, \href
  {http://adsabs.harvard.edu/abs/2010ApJ...715..833O} {715, 833}

\bibitem[\protect\citeauthoryear{{Pesce} et~al.,}{{Pesce}
  et~al.}{2020}]{pesce2020}
{Pesce} D.~W.,  et~al., 2020, \mn@doi [\apjl] {10.3847/2041-8213/ab75f0}, \href
  {https://ui.adsabs.harvard.edu/abs/2020ApJ...891L...1P} {891, L1}

\bibitem[\protect\citeauthoryear{{Pietrzy{\'n}ski} et~al.,}{{Pietrzy{\'n}ski}
  et~al.}{2019}]{pietrzynski19}
{Pietrzy{\'n}ski} G.,  et~al., 2019, \mn@doi [\nat]
  {10.1038/s41586-019-0999-4}, \href
  {https://ui.adsabs.harvard.edu/abs/2019Natur.567..200P} {567, 200}

\bibitem[\protect\citeauthoryear{{Planck Collaboration} et~al.,}{{Planck
  Collaboration} et~al.}{2018}]{planck18}
{Planck Collaboration} et~al., 2018, arXiv e-prints, \href
  {https://ui.adsabs.harvard.edu/abs/2018arXiv180706209P} {p. arXiv:1807.06209}

\bibitem[\protect\citeauthoryear{{Polshaw} et~al.,}{{Polshaw}
  et~al.}{2015}]{polshaw15}
{Polshaw} J.,  et~al., 2015, \mn@doi [\aap] {10.1051/0004-6361/201526902},
  \href {https://ui.adsabs.harvard.edu/abs/2015A&A...580L..15P} {580, L15}

\bibitem[\protect\citeauthoryear{{Poznanski}, {Butler}, {Filippenko}
  et~al.}{{Poznanski} et~al.}{2009}]{poznanski09}
{Poznanski} D.,  {Butler} N.,  {Filippenko} A.~V.,   et~al., 2009, \mn@doi
  [\apj] {10.1088/0004-637X/694/2/1067}, \href
  {http://adsabs.harvard.edu/abs/2009ApJ...694.1067P} {694, 1067}

\bibitem[\protect\citeauthoryear{{Reid}, {Pesce}  \& {Riess}}{{Reid}
  et~al.}{2019}]{reid19}
{Reid} M.~J.,  {Pesce} D.~W.,   {Riess} A.~G.,  2019, \mn@doi [\apjl]
  {10.3847/2041-8213/ab552d}, \href
  {https://ui.adsabs.harvard.edu/abs/2019ApJ...886L..27R} {886, L27}

\bibitem[\protect\citeauthoryear{{Riess} et~al.,}{{Riess}
  et~al.}{2009}]{riess09}
{Riess} A.~G.,  et~al., 2009, \mn@doi [\apj] {10.1088/0004-637X/699/1/539},
  \href {https://ui.adsabs.harvard.edu/abs/2009ApJ...699..539R} {699, 539}

\bibitem[\protect\citeauthoryear{{Riess}, {Macri}, {Casertano}  et~al.}{{Riess}
  et~al.}{2011}]{riess11}
{Riess} A.~G.,  {Macri} L.,  {Casertano} S.,   et~al., 2011, \mn@doi [\apj]
  {10.1088/0004-637X/730/2/119}, \href
  {http://adsabs.harvard.edu/abs/2011ApJ...730..119R} {730, 119}

\bibitem[\protect\citeauthoryear{{Riess}, {Macri}, {Hoffmann}  et~al.}{{Riess}
  et~al.}{2016}]{riess16}
{Riess} A.~G.,  {Macri} L.~M.,  {Hoffmann} S.~L.,   et~al., 2016, \mn@doi
  [\apj] {10.3847/0004-637X/826/1/56}, \href
  {http://adsabs.harvard.edu/abs/2016ApJ...826...56R} {826, 56}

\bibitem[\protect\citeauthoryear{{Riess}, {Casertano}, {Yuan}  et~al.}{{Riess}
  et~al.}{2018a}]{riess18a}
{Riess} A.~G.,  {Casertano} S.,  {Yuan} W.,   et~al., 2018a, \mn@doi [\apj]
  {10.3847/1538-4357/aaadb7}, \href
  {http://adsabs.harvard.edu/abs/2018ApJ...855..136R} {855, 136}

\bibitem[\protect\citeauthoryear{{Riess} et~al.,}{{Riess}
  et~al.}{2018b}]{riess18b}
{Riess} A.~G.,  et~al., 2018b, \mn@doi [\apj] {10.3847/1538-4357/aac82e}, \href
  {https://ui.adsabs.harvard.edu/abs/2018ApJ...861..126R} {861, 126}

\bibitem[\protect\citeauthoryear{{Riess}, {Casertano}, {Yuan}, {Macri}  \&
  {Scolnic}}{{Riess} et~al.}{2019}]{riess19}
{Riess} A.~G.,  {Casertano} S.,  {Yuan} W.,  {Macri} L.~M.,   {Scolnic} D.,
  2019, \mn@doi [\apj] {10.3847/1538-4357/ab1422}, \href
  {https://ui.adsabs.harvard.edu/abs/2019ApJ...876...85R} {876, 85}

\bibitem[\protect\citeauthoryear{{Riess} et~al.,}{{Riess}
  et~al.}{2021a}]{riess22}
{Riess} A.~G.,  et~al., 2021a, arXiv e-prints, \href
  {https://ui.adsabs.harvard.edu/abs/2021arXiv211204510R} {p. arXiv:2112.04510}

\bibitem[\protect\citeauthoryear{{Riess}, {Casertano}, {Yuan}, {Bowers},
  {Macri}, {Zinn}  \& {Scolnic}}{{Riess} et~al.}{2021b}]{riess21}
{Riess} A.~G.,  {Casertano} S.,  {Yuan} W.,  {Bowers} J.~B.,  {Macri} L.,
  {Zinn} J.~C.,   {Scolnic} D.,  2021b, \mn@doi [\apjl]
  {10.3847/2041-8213/abdbaf}, \href
  {https://ui.adsabs.harvard.edu/abs/2021ApJ...908L...6R} {908, L6}

\bibitem[\protect\citeauthoryear{{Rodr{\'{\i}}guez} et~al.,}{{Rodr{\'{\i}}guez}
  et~al.}{2019}]{rodriguez19a}
{Rodr{\'{\i}}guez} {\'O}.,  et~al., 2019, \mn@doi [\mnras]
  {10.1093/mnras/sty3396}, \href
  {http://adsabs.harvard.edu/abs/2019MNRAS.483.5459R} {483, 5459}

\bibitem[\protect\citeauthoryear{{Sandage}, {Tammann}, {Saha}, {Reindl},
  {Macchetto}  \& {Panagia}}{{Sandage} et~al.}{2006}]{sandage06}
{Sandage} A.,  {Tammann} G.~A.,  {Saha} A.,  {Reindl} B.,  {Macchetto} F.~D.,
  {Panagia} N.,  2006, \mn@doi [\apj] {10.1086/508853}, \href
  {https://ui.adsabs.harvard.edu/abs/2006ApJ...653..843S} {653, 843}

\bibitem[\protect\citeauthoryear{{Schmidt}, {Kirshner}, {Eastman}
  et~al.}{{Schmidt} et~al.}{1994}]{schmidt94}
{Schmidt} B.~P.,  {Kirshner} R.~P.,  {Eastman} R.~G.,   et~al., 1994, \mn@doi
  [\apj] {10.1086/174546}, \href
  {http://adsabs.harvard.edu/abs/1994ApJ...432...42S} {432, 42}

\bibitem[\protect\citeauthoryear{{Sedgwick}, {Collins}, {Baldry}  \&
  {James}}{{Sedgwick} et~al.}{2021}]{sedgwick21}
{Sedgwick} T.~M.,  {Collins} C.~A.,  {Baldry} I.~K.,   {James} P.~A.,  2021,
  \mn@doi [\mnras] {10.1093/mnras/staa3456}, \href
  {https://ui.adsabs.harvard.edu/abs/2021MNRAS.500.3728S} {500, 3728}

\bibitem[\protect\citeauthoryear{{Spergel}, {Bean}, {Dor{\'e}}
  et~al.}{{Spergel} et~al.}{2007}]{spergel07}
{Spergel} D.~N.,  {Bean} R.,  {Dor{\'e}} O.,   et~al., 2007, \mn@doi [\apjs]
  {10.1086/513700}, \href {http://adsabs.harvard.edu/abs/2007ApJS..170..377S}
  {170, 377}

\bibitem[\protect\citeauthoryear{{Van Dyk} et~al.,}{{Van Dyk}
  et~al.}{2019}]{vandyk19}
{Van Dyk} S.~D.,  et~al., 2019, arXiv e-prints, \href
  {http://adsabs.harvard.edu/abs/2019arXiv190303872V} {}

\bibitem[\protect\citeauthoryear{{Virtanen} et~al.,}{{Virtanen}
  et~al.}{2020}]{scipy}
{Virtanen} P.,  et~al., 2020, \mn@doi [Nature Methods]
  {10.1038/s41592-019-0686-2}, \href
  {https://ui.adsabs.harvard.edu/abs/2020NatMe..17..261V} {17, 261}

\bibitem[\protect\citeauthoryear{{Vogl}}{{Vogl}}{2020}]{voglphd}
{Vogl} C.,  2020, PhD thesis, Technical University of Munich, Germany

\bibitem[\protect\citeauthoryear{{Vogl}, {Sim}, {Noebauer}, {Kerzendorf}  \&
  {Hillebrandt}}{{Vogl} et~al.}{2019}]{vogl19}
{Vogl} C.,  {Sim} S.~A.,  {Noebauer} U.~M.,  {Kerzendorf} W.~E.,
  {Hillebrandt} W.,  2019, \mn@doi [\aap] {10.1051/0004-6361/201833701}, \href
  {https://ui.adsabs.harvard.edu/abs/2019A&A...621A..29V} {621, A29}

\bibitem[\protect\citeauthoryear{{Vogl}, {Kerzendorf}, {Sim}, {Noebauer},
  {Lietzau}  \& {Hillebrandt}}{{Vogl} et~al.}{2020}]{vogl20}
{Vogl} C.,  {Kerzendorf} W.~E.,  {Sim} S.~A.,  {Noebauer} U.~M.,  {Lietzau} S.,
    {Hillebrandt} W.,  2020, \mn@doi [\aap] {10.1051/0004-6361/201936137},
  \href {https://ui.adsabs.harvard.edu/abs/2020A&A...633A..88V} {633, A88}

\bibitem[\protect\citeauthoryear{{Whitelock}, {Feast}  \& {Van
  Leeuwen}}{{Whitelock} et~al.}{2008}]{whitelock08}
{Whitelock} P.~A.,  {Feast} M.~W.,   {Van Leeuwen} F.,  2008, \mn@doi [\mnras]
  {10.1111/j.1365-2966.2008.13032.x}, \href
  {https://ui.adsabs.harvard.edu/abs/2008MNRAS.386..313W} {386, 313}

\bibitem[\protect\citeauthoryear{{Yuan}, {Riess}, {Macri}, {Casertano}  \&
  {Scolnic}}{{Yuan} et~al.}{2019}]{yuan2019}
{Yuan} W.,  {Riess} A.~G.,  {Macri} L.~M.,  {Casertano} S.,   {Scolnic} D.~M.,
  2019, \mn@doi [\apj] {10.3847/1538-4357/ab4bc9}, \href
  {https://ui.adsabs.harvard.edu/abs/2019ApJ...886...61Y} {886, 61}

\bibitem[\protect\citeauthoryear{{Yuan} et~al.,}{{Yuan} et~al.}{2020}]{yuan20}
{Yuan} W.,  et~al., 2020, \mn@doi [\apj] {10.3847/1538-4357/abb377}, \href
  {https://ui.adsabs.harvard.edu/abs/2020ApJ...902...26Y} {902, 26}

\makeatother
\end{thebibliography}
\end{document}